\documentclass{article}

\usepackage{PRIMEarxiv}
\usepackage[toc,page,title]{appendix} 
\usepackage[utf8]{inputenc} 
\usepackage[T1]{fontenc}    
\usepackage{hyperref}       
\usepackage{url}            
\usepackage{booktabs}       
\usepackage{tabularx}       
\usepackage{physics}        
\usepackage{amsfonts}       
\usepackage{subcaption}
\usepackage{nicefrac}       
\usepackage{tikz}           
\usepackage{microtype}      
\usepackage{amsthm}         
\newtheorem{lemma}{Lemma}
\newtheorem{theorem}{Theorem}
\newtheorem{corollary}{Corollary}
\usepackage{booktabs,amsmath,bm}
\usepackage{pgfplots}\pgfplotsset{compat=1.17}
\usepgfplotslibrary{groupplots}

\usepackage{tikz}
\usepackage{graphicx}
\usetikzlibrary{arrows.meta, positioning, calc}

\usepackage{amssymb}        

\usepackage{lipsum}
\usepackage{amsmath}
\usepackage{natbib}
\usepackage{xcolor}
\usepackage{algorithm,algpseudocode}
\usepackage{fancyhdr}       
\usepackage{graphicx}       
\graphicspath{{media/}}     
\usepackage[normalem]{ulem}

\title{Max-Cut QAOA on Non-Hardware-Native Graphs via MinLA Surrogates and Configuration Recovery
}

\author{
  K. Bharati \\
  Indian Institute of Technology Madras \thanks{\textit{K. Bharati worked on this research project during her summer internship at IBM Research.}} \\
  \texttt{kbharati1998@gmail.com} \\
   \And
  Nathan Earnest-Noble \\
  IBM Research \\
  \texttt{nate@ibm.com} \\
   \And
   Ritajit Majumdar \\
   IBM Research \\
   \texttt{majumdar.ritajit@ibm.com} \\
}

\begin{document}
\maketitle



\begin{abstract}

    For non-hardware-native \textsc{MAX-CUT} instances, the Quantum Approximate Optimization Algorithm (QAOA) can incur additional two-qubit depth overhead due to SWAP-based routing of interactions between non-adjacent qubits. 
    We propose a hybrid quantum--classical workflow that reduces this depth by selectively removing interactions and subsequently recovers solution quality through classical optimization of the original \textsc{MAX-CUT} objective.
    We employ a Minimum Linear Arrangement (MinLA)-based vertex ordering to preferentially retain short-range interactions, thereby constructing a two-qubit depth-reduced surrogate QAOA ansatz.
    %
    To compensate for the information lost through interaction removal, we develop classical \textit{recovery} strategies that refine QPU-sampled bitstrings directly on the original \textsc{MAX-CUT} objective.
    We develop two recovery strategies: signed-residue-based local search and bipartite-block optimization, where each conditional block problem is solved exactly using an $s$--$t$ minimum cut. Recovery is performed independently on each sampled configuration, ensuring monotonic improvement in cut value without additional QPU executions.
    We evaluate the workflow on \texttt{ibm\_kingston} for 120 distinct non-hardware-native instances comprising $6$-regular and Erd\H{o}s--R\'enyi graphs with 40, 60 and 100 vertices. 
    We verify that the surrogates substantially reduce transpiled two-qubit depth while staying within the prescribed two-qubit depth, and with recovery yield \textit{near-optimal} cuts on the original graphs. 
    %
    These results demonstrate the potential of hardware-aware interaction selection and classical recovery to extend QAOA to graph instances whose full cost Hamiltonians exceed practical circuit-depth budgets.
\end{abstract}

\keywords{Quantum Approximate Optimization Algorithm (QAOA) \and MAX-CUT \and Minimum Linear Arrangement (MinLA) \and Hardware-Aware Quantum Optimization \and Surrogate Hamiltonians \and Configuration Recovery}


\section{Introduction}
\label{sec:introduction}

Many combinatorial optimization problems admist natural graph-based formulations, in which the objective is to identify a configuration that optimizes a function defined over the graph. 
%
%
The Quantum Approximate Optimization Algorithm (QAOA) \cite{farhi2014quantum} is a hybrid quantum--classical algorithm for \textit{approximately} solving such problems, with particular relevance to near-term quantum hardware. 
%
%
%
QAOA prepares parameterized quantum states through alternating evolutions under a problem-specific \textit{cost Hamiltonian} and a \textit{mixer Hamiltonian}, with the parameters optimized through a classical feedback loop. 
In this work, we focus on \textsc{MAX-CUT}, an NP-hard and APX-hard problem that seeks a vertex partition into two disjoint sets so as to maximize the number of edges crossing between them.

\begin{figure*}[t]
\centering
\includegraphics[width=\textwidth]{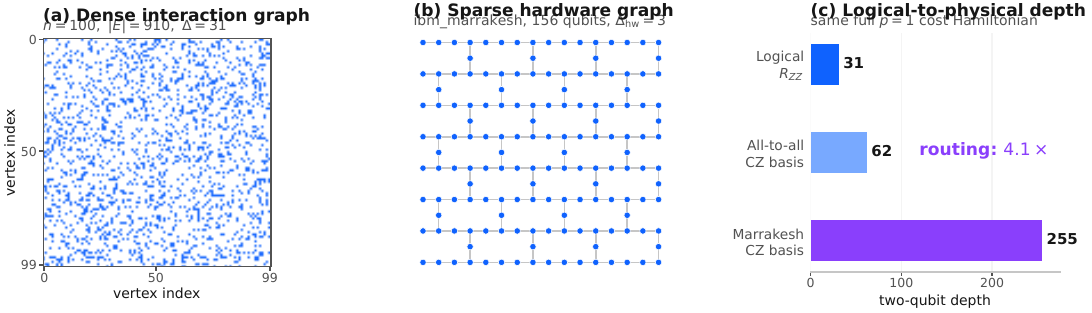}
\caption{Connectivity-induced depth inflation for the $p=1$ QAOA cost Hamiltonian circuit of the HAMLIB instance \texttt{gnp-k\_4\_n-100\_rinst-00}. Although the 910 edge interactions require a two-qubit depth of only 62, implementing the same circuit on the sparse coupling map of \texttt{ibm\_kingston} increases the depth to 255 because of the additional SWAP operations needed to route non-local interactions. The resulting increase in noise can negate the benefit of executing the full problem Hamiltonian, motivating hardware-aware reductions that preserve useful problem structure while remaining executable on current superconducting processors.}

\label{fig:depth-motivation}
\end{figure*}

{For \textsc{MAX-CUT}, each graph edge corresponds to a two-qubit interaction in the QAOA cost Hamiltonian. As a result, the structure of the input graph directly determines the interactions that must be executed on quantum hardware.} So \textsc{MAX-CUT} directly exposes the mismatch between a dense problem graph and the sparse coupling map of a superconducting quantum processor. Interactions between uncoupled qubits must be routed with SWAP gates, inflating the two-qubit depth~\cite{Harrigan_2021,Weidenfeller2022}. {For the 100-vertex, 910-edge graph of Fig.~\ref{fig:depth-motivation}, the QAOA cost layer itself requires a two-qubit depth of only 62. However, mapping the circuit onto the sparse coupling map of \texttt{ibm\_kingston} requires many additional SWAP gates to realize non-local interactions, increasing the depth to 255. The resulting increase in circuit depth amplifies the effect of hardware noise, often overwhelming any benefit obtained from executing the full problem Hamiltonian.} 



Earlier experimental studies on MAX-CUT investigated graphs with up to 23 and 40 vertices~\cite{Harrigan_2021,SackEgger2024}. More recent experiments involving over 100 qubits have explored two contrasting approaches. In~\cite{Sachdeva2026}, the authors execute a standard $p=1$ QAOA cost layer, incorporating error suppression and greedy bit-flip post-processing, on 3- and 5-regular graphs with up to 156 vertices. These instances remain relatively sparse, with the largest containing 390 edges. In contrast,~\cite{wang2025depth} reduces circuit depth by retaining only interactions along a single path through the graph, limiting the executed Hamiltonian to at most $n-1$ interactions.
Dense graphs present a different challenge. As illustrated in Fig.~\ref{fig:depth-motivation}, executing the full cost Hamiltonian can require excessive two-qubit depth, while restricting interactions to a single path may discard more of the original graph structure than necessary. In particular, current superconducting hardware may support circuits containing substantially more than $n-1$ interactions within a feasible depth budget. The objective is therefore to retain as much of the original graph structure as possible without exceeding the circuit-depth budget of the target hardware.

On sparsely connected quantum processors, non-local interactions can require costly routing, whereas their contributions to a cut can be evaluated classically at low cost. We exploit this asymmetry by reordering graph vertices to favor shorter-range interactions and selecting a subset for quantum execution based on the resulting transpiled circuit depth. Unlike hardware-aware compilation~\cite{Weidenfeller2022,MATSUO_2023} and graph sparsification with cut-approximation guarantees~\cite{Moondra_2026}, our approach explicitly uses the transpiled depth on the target device to guide interaction selection. However, removing interactions modifies the cost Hamiltonian, potentially changing its optimal solutions. We therefore refine the sampled configurations classically against the original \textsc{MAX-CUT} objective. This motivates our central question: \emph{how can we select interactions under a hardware depth budget while retaining sufficient problem structure to recover high-quality solutions to the original problem?}



\begin{figure}[htbp]
\centering
\begin{tikzpicture}[
  node distance=4mm,
  font=\footnotesize,
  box/.style={
    draw, rounded corners=2pt,
    text width=9cm,
    minimum height=9mm,
    align=center,
    inner sep=4pt
  },
  neutral/.style={box,fill=gray!8},
  classical/.style={
    box,fill=green!9,draw=green!50!black
  },
  quantum/.style={
    box,fill=blue!9,draw=blue!60!black
  },
  arr/.style={-{Latex[length=2mm]},thick}
]
\node[neutral] (a) {
  \textbf{Original \textsc{MAX-CUT} Problem}\\
  Graph $G$
};

\node[classical,below=of a] (b) {
  \textbf{Hardware-Native Surrogate Construction}\\
  Select interactions within the hardware depth budget
};

\node[quantum,below=of b] (c) {
  \textbf{QAOA Execution on the Surrogate Hamiltonian}\\
  Quantum circuit execution and bitstring sampling
};
\node[classical,below=of c] (d) {
  \textbf{Recover Solution Quality in Parallel}\\
  Independent refinement of top-$B$ bitstrings on the original graph
};
\node[neutral,below=of d] (e) {
  \textbf{Select the Best Recovered Solution}\\
  Highest cut value on the original \textsc{MAX-CUT} graph
};
\draw[arr] (a) -- (b);
\draw[arr] (b) -- (c);
\draw[arr] (c) -- (d);
\draw[arr] (d) -- (e);
\end{tikzpicture}
\caption{Principle of the proposed hybrid quantum--classical
workflow. A depth-constrained surrogate reduces the quantum
workload, while the top-$B$ sampled configurations are refined
independently and in parallel against the original
\textsc{MAX-CUT} objective. Green and blue denote classical
and quantum processing, respectively.}
\label{fig:workflow}
\end{figure}
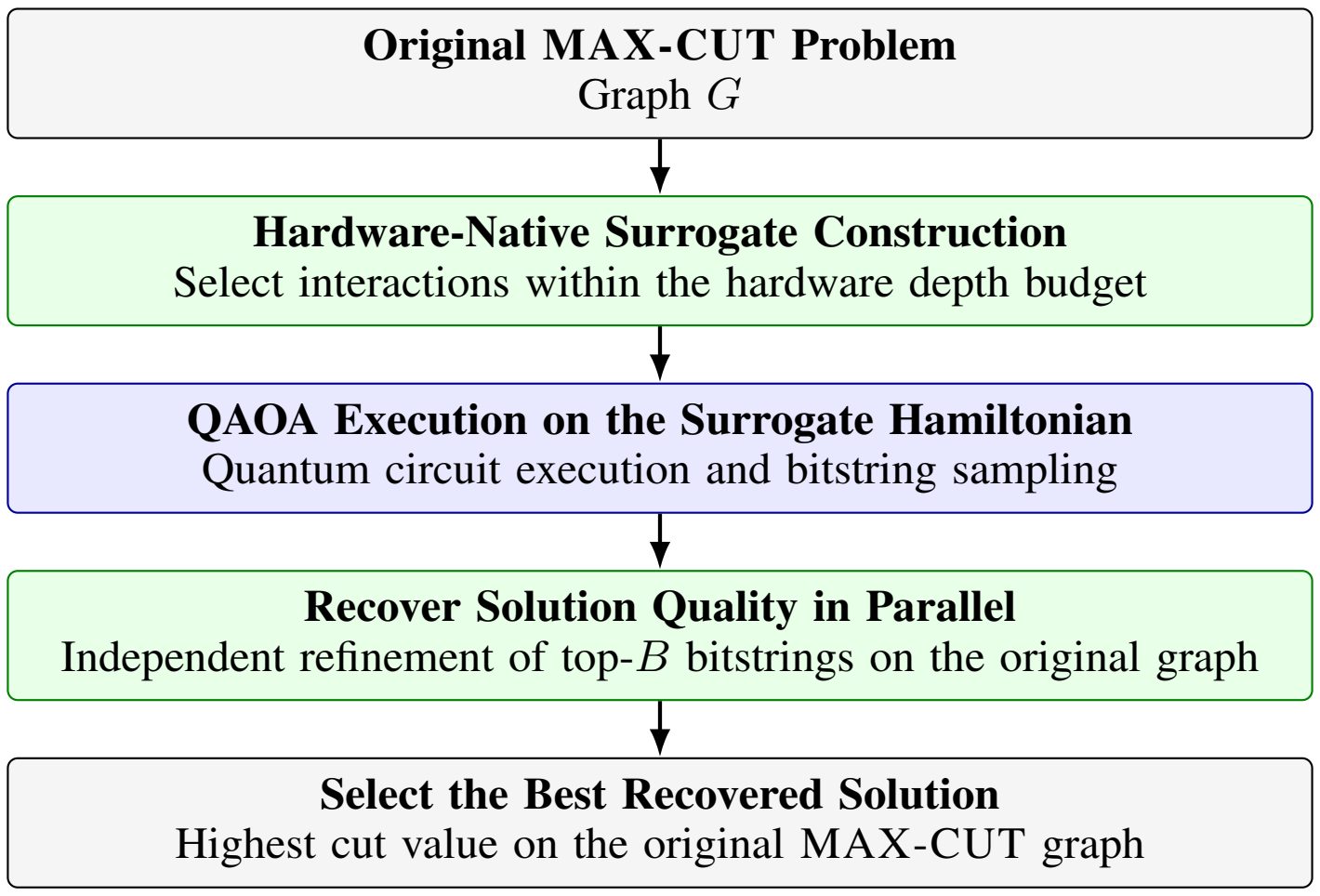

We propose a hybrid quantum--classical workflow (Fig.~\ref{fig:workflow}) that prioritizes quantum circuit depth while preserving the ability to obtain high-quality solutions to the original \textsc{MAX-CUT} problem. The workflow first constructs a surrogate Hamiltonian by selecting a subset of interactions that can be executed within a prescribed depth budget. QAOA is then executed using this surrogate Hamiltonian to generate candidate configurations. For the graph instances considered in this work, QAOA angles are selected using existing classical methods~\cite{qaoa-training-pipeline}, avoiding iterative QPU-based parameter training. However, this is an experimental choice rather than a requirement of the framework, which remains compatible with QPU-based variational optimization, as considered in ~\cite{Sachdeva2026}. The top-$B$ sampled configurations are subsequently refined independently and in parallel through classical optimization of the original \textsc{MAX-CUT} objective, and the best recovered solution is returned.

%

The main contributions of this paper are enlisted below:
\begin{enumerate}
    \item \textit{Hardware-native surrogate construction:} We reorder vertices using minimum linear arrangement (MinLA) to concentrate interactions between nearby vertices and retain a diagonal band of interactions in the reordered graph. The resulting surrogate Hamiltonian preserves substantially more of the original problem structure than a chain while remaining compatible with hardware connectivity constraints. The construction interpolates between the two extremes above: under a path-following ordering, a band of radius one reduces to a chain, while a band as wide as the bandwidth recovers the full cost Hamiltonian. Every executed interaction belongs to the original Hamiltonian, and every decision variable is retained.

    \item \textit{Depth-budgeted surrogate selection:} We choose the widest executable interaction band whose transpiled $p$-layer circuit remains below a device-specific depth budget $L$, while ensuring that every vertex retains at least one incident interaction. This directly couples surrogate-graph construction to the capabilities of the target hardware.

    \item \textit{Configuration Recovery for MAX-CUT:} Motivated by configuration recovery mechanism in sample-based quantum diagonalization (SQD) ~\cite{RobledoMoreno2025}, we propose two configuration recovery algorithms that operate directly on discrete assignments rather than through subspace diagonalization. 
    \begin{enumerate}
         \item \textit{Greedy Signed-residue recovery:} First, we propose a greedy procedure that flips single vertices whose signed residue on the full graph is positive, increasing the cut monotonically.
    
        \item \textit{Bipartite-block recovery:} To circumvent the local optima of single-vertex flips, we propose a multi-vertex update policy that re-optimizes the union of two colour classes with the rest of the assignment fixed. The bipartite structure of each block reduces this conditional problem to an $s$--$t$ minimum cut~\cite{kolmogorov2004energy}.
    \end{enumerate}

    \item \textit{Parallel recovery from a single QPU execution:} For statistical accuracy, we perform independent recovery runs on each of the top-$B$ sampled QPU bitstrings. These runs share no state, so they execute in parallel on classical processors, and the best recovered cut is returned without further QPU calls.

    \item \textit{Experimental Evaluation on Quantum Hardware:} We evaluate the workflow on a combined total of 120 non-hardware-native 6-regular and Erd\H{o}s--R\'enyi ($G(n,p)$) instances from HAMLIB~\cite{sawaya2024hamlib}, with up to 100 vertices, using the 156-qubit IBM Heron r2 processor \texttt{ibm\_kingston}. Our results demonstrate that selecting interactions under a hardware-specific circuit-depth budget allows reduced QAOA circuits to retain more of the original graph structure than chain-based reductions. Combined with classical recovery, this enables high-quality solutions to be obtained for the original \textsc{MAX-CUT} instances without executing their full cost Hamiltonians. 
\end{enumerate}

The proposed workflow illustrated in Fig.~\ref{fig:workflow} naturally aligns with futuristic hardware improvements: as lower-noise processors support deeper circuits, increasingly larger interaction bands can be executed, allowing surrogate Hamiltonians to capture more of the original problem structure and extending the range of non-hardware-native optimization problems that can be addressed.

The remainder of the paper is organized as follows. Section~\ref{sec:prelim} reviews the QAOA formulation, parameter-selection strategy, and hardware-aware compilation methods used in this work. Section~\ref{sec:interaction-filtering} presents the construction of hardware-feasible surrogate Hamiltonians through MinLA-based vertex ordering and depth-constrained interaction filtering. Section~\ref{sec:recovery} introduces the proposed configuration-recovery procedures for restoring information omitted during quantum execution. Section~\ref{subsec:experimental-protocol} evaluates the complete workflow on non-hardware-native \textsc{MAX-CUT} instances using IBM quantum hardware, and Section~\ref{sec:conclusion} concludes with a discussion of implications and future directions.



\section{Background and Preliminaries}
\label{sec:prelim}

We consider an undirected, unweighted graph
$G=(V,E)$ with $|V|=n$ vertices. The
\textsc{MAX-CUT} problem seeks a bipartition
of the vertices that maximizes the number of
edges crossing the partition. Assigning a
binary variable $x_i\in\{0,1\}$ to each
vertex, the cut objective is
\begin{equation}
\label{eq:maxcut-qubo}
C_G(\boldsymbol{x})
=
\sum_{(i,j)\in E}
(x_i+x_j-2x_ix_j).
\end{equation}
Using $x_i=(1-z_i)/2$, where
$z_i\in\{-1,+1\}$, and replacing each spin
variable with the Pauli-$Z$ operator, the
corresponding cost Hamiltonian is
\begin{equation}
\label{eq:maxcut-ising}
H_C =
\frac{1}{2}\sum_{(i,j)\in E}
(I-Z_iZ_j).
\end{equation}

The Quantum Approximate Optimization Algorithm
(QAOA)~\cite{farhi2014quantum} alternates
evolution under the cost Hamiltonian $H_C$
and the mixer Hamiltonian
$H_M=\sum_{i\in V}X_i$.
The $p$-layer QAOA state is
\begin{equation}
\label{eq:qaoa-state}
\begin{aligned}
\ket{\psi_p(\boldsymbol{\theta})}
=
U_M(\beta_p)U_C(\gamma_p)\cdots
&U_M(\beta_1)U_C(\gamma_1)
\ket{+}^{\otimes n},
\end{aligned}
\end{equation}
where $U_C(\gamma)=e^{-i\gamma H_C}$,
$U_M(\beta)=e^{-i\beta H_M}$, and
$\boldsymbol{\theta}=
(\boldsymbol{\gamma},\boldsymbol{\beta})$
denotes the variational parameters. Each cost layer comprises $|E|$ commuting
two-qubit interactions, whose implementation
depth depends on the hardware connectivity
and compilation strategy. Further details
on the QAOA formulation, parameter-selection
methods, and connectivity-aware compilation
are detailed in Appendix \ref{app:prelim}.

\section{Depth-Constrained Interaction Filtering}
\label{sec:interaction-filtering}

The techniques described in Appendix~\ref{subsec:circuit-compilation} reduce routing overhead during circuit compilation. However, for dense graphs, compilation alone may not be sufficient to make QAOA circuits practical on current hardware. As illustrated in Fig.~\ref{fig:depth-motivation}, routing full problem Hamiltonians can increase the two-qubit depth many folds, leading to a degradation in the quality of the sampled solutions. To improve the hardware feasibility of such instances, we reduce the interaction set before circuit synthesis and modify the optimisation Hamiltonian.


Graph reordering provides a natural basis for such interaction reduction. Classical graph-layout methods, including bandwidth minimization and Minimum Linear Arrangement (MinLA), seek vertex orderings that reduce the separation between interacting vertices ~\citep{diaz2002survey}. Efficient heuristics, including multilevel approaches, have been developed for obtaining low-cost linear arrangements~\citep{safro2006}.In the context of QAOA, graph sparsification has also been investigated to reduce implementation complexity while approximately preserving cut values ~\citep{Moondra_2026}.

Our choice of MinLA is motivated by the path-based routing strategies employed on the near-linear hardware coupling graph of Heron architectures \cite{Weidenfeller2022}. In such arrangements, interactions between distant vertices can incur additional routing overhead. By minimizing the cumulative separation between interacting vertices, MinLA encourages the concentration of non-zero entries along the first few superdiagonals of the reordered adjacency matrix, providing a natural basis for identifying short-range interactions. This allows a larger fraction of the original graph to be retained within a fixed interaction radius, producing a more informative surrogate Hamiltonian under the same hardware depth budget.

In this section, we describe how we exploit this structure to construct a band-limited interaction graph, preferentially retaining interactions near the diagonal while discarding longer-range ones. Unlike conventional graph reordering, which preserves all edges, our approach explicitly reduces the interaction set. Rather than imposing a fixed sparsification ratio or a cut-approximation constraint, we select the largest interaction radius satisfying two requirements: every vertex must retain at least one incident edge, and the transpiled two-qubit depth of the resulting $p$-layer QAOA circuit must remain below a prescribed hardware depth budget $L$. This budget reflects the practical requirement that circuit depth remain low enough to limit the impact of accumulated hardware noise.


\subsection{MinLA-Based Interaction Ordering}
\label{subsec:minla-ordering}

The objective of the proposed workflow is not merely to reduce the number of interactions, but to retain as much of the original graph structure as possible subject to a hardware depth budget. To achieve this, we seek a vertex ordering that places many interacting vertex pairs close together. Such an ordering allows subsequent interaction filtering to preserve a larger fraction of the original graph while maintaining a low circuit depth. This motivates our use of the Minimum Linear Arrangement (MinLA) problem.

Let $\mathbf{A}$ denote the adjacency matrix of the graph $G=(V,E)$. A linear arrangement is defined as a bijection
$\pi:V\rightarrow\{0,\ldots,n-1\}$ that assigns
each vertex a distinct position. The MinLA
problem identifies an ordering that minimizes the
cumulative distance between adjacent vertices,
    \begin{equation}  \label{eq:minla-objective}
        \min_{\pi} 
        \sum_{(u,v)\in E}|\pi(u)-\pi(v)|.
    \end{equation}

Unlike bandwidth minimization strategies like Reverse Cuthill-Mckee \cite{liu1976, CuthillMcKee1969} used in the context of QAOA \cite{dasgupta2026quantum}, which penalizes the largest individual edge span, MinLA minimizes the sum of all edge spans. Since finding an optimal arrangement is NP-hard, we implement the heuristic proposed in \cite{safro2006} to obtain low-cost orderings. Let $\pi_{\mathrm{LA}}$ denote the MinLA
ordering. The corresponding adjacency matrix
is obtained by applying the same permutation
to its rows and columns,
    \begin{equation} \label{eq:reordered-adjacency}
        (\mathbf{A}_{\pi})_{ij}
        =\mathbf{A}_{\pi_{\mathrm{LA}}^{-1}(i),
        \pi_{\mathrm{LA}}^{-1}(j)}.
    \end{equation}

\begin{figure}[htbp]
    \centering
    \includegraphics[width=\linewidth]{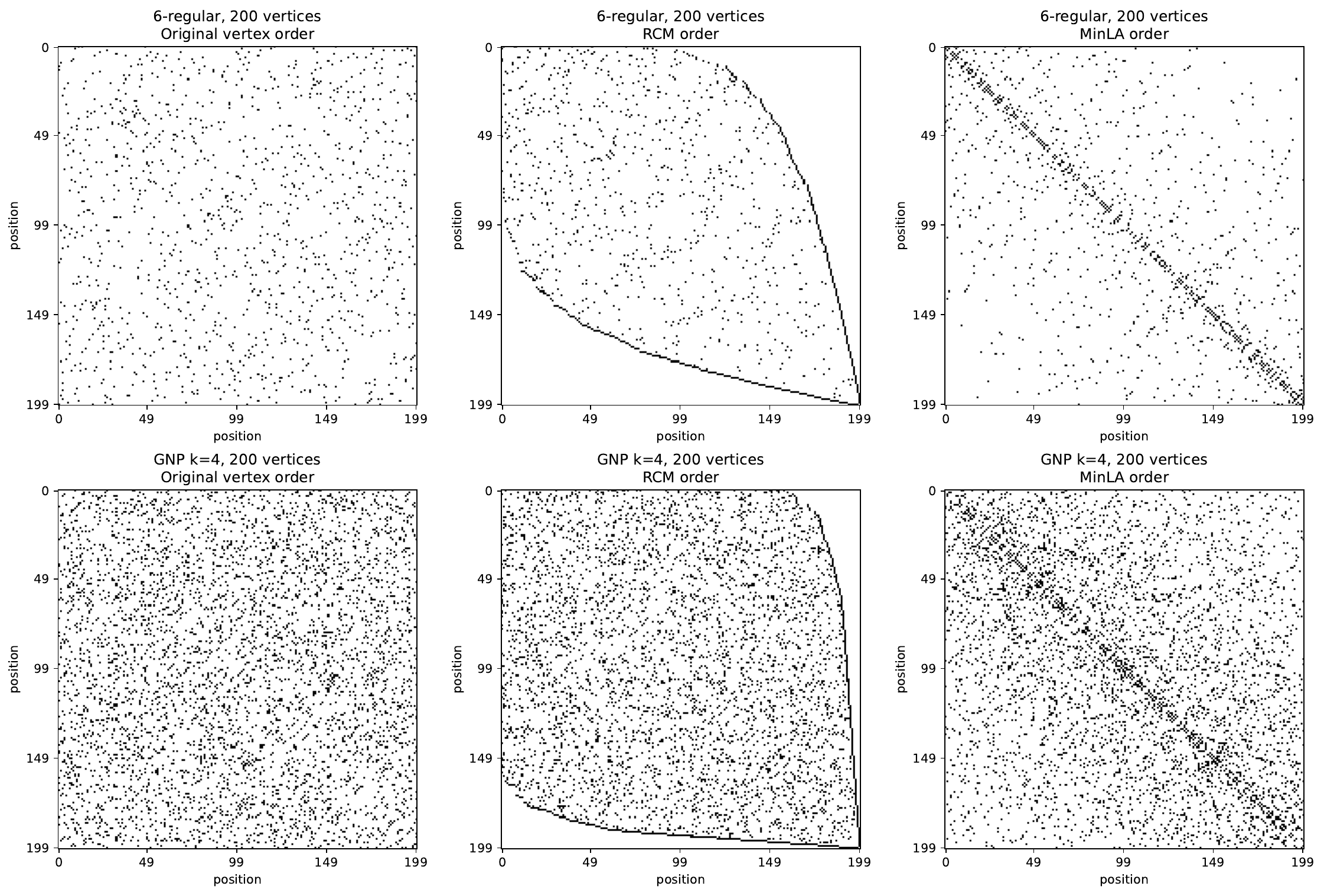}
    \caption{Original, RCM and MinLA vertex ordering of HAMLIB instances of (a) a 6-regular graph and (b) a $G(n,p)$ graph.}
    \label{fig:minla-filtering}
\end{figure}

As mentioned previously, the superdiagonals of $\mathbf{A}_{\pi}$ provide a natural representation of interaction span. An entry on the $r$th superdiagonal represents an edge whose endpoints are separated by $r$ positions in the linear arrangement. Denoting the number of nonzero entries on this superdiagonal by $s_r$, the arrangement cost is $\sum_{r=1}^{n-1}r\,s_r$. Minimizing this quantity encourages the
concentration of interactions along the first
few superdiagonals, thereby favoring shorter
interaction spans. This property motivates
restricting the QAOA cost Hamiltonian to
interactions lying within a narrow diagonal
band, as illustrated in Fig.~\ref{fig:minla-filtering}. The adjacency matrices of the 200-vertex HAMLIB 6-regular and $G(n,p)$ graphs \cite{sawaya2024hamlib} are displayed in their original, RCM, and MinLA vertex orders. We can clearly observe that MinLA places more edge entries close to the main diagonal than RCM, especially at short spans. Consequently, a larger fraction of the original interactions can be preserved when only a limited interaction radius is retained.

\subsection{Depth-Constrained Selection of Interactions}
\label{subsec:depth-constrained-selection}

For a given ordering $\pi_{\mathrm{LA}}$, we
construct a reduced interaction graph
$G_k=(V,E_k)$ by retaining edges whose spans
do not exceed an integer radius $k$,
    \begin{equation} \label{eq:retained-interactions}
        E_k= \left\{(u,v)\in E:
        |\pi_{\mathrm{LA}}(u)-
        \pi_{\mathrm{LA}}(v)|\leq k
        \right\}.
    \end{equation}
Equivalently, $E_k$ contains the original interactions represented within the first $k$ superdiagonals and their symmetric subdiagonals. Since $E_k\subseteq E_{k+1}$, increasing $k$ progressively restores interactions from
the original graph. The choice of $k$ determines the tradeoff between information retention and hardware cost: larger values preserve more interactions from the original graph, but generally increase routing overhead and circuit depth. An excessively narrow
band may isolate vertices, leaving their associated variables absent from the surrogate cost Hamiltonian. To avoid this, we mandate \textit{every vertex to retain at least one incident edge}. Assuming that $G$ contains no isolated vertices, the smallest radius satisfying
this condition is
    \begin{equation} \label{eq:coverage-radius}
        k_{\mathrm{cov}}=
        \min\left\{
        k:\deg_{G_k}(v)\geq1,\quad
        \forall v\in V
        \right\}.
    \end{equation}
Equivalently, \eqref{eq:coverage-radius} can be expressed as,
    \begin{equation}
    k_{\mathrm{cov}}=
    \max_{v\in V}
    \min_{u\in N_G(v)}
    |\pi_{\mathrm{LA}}(u)-
    \pi_{\mathrm{LA}}(v)|,
    \label{eq:coverage-equivalent}
    \end{equation}
where $N_G(v)$ denotes the neighborhood of
vertex $v$ in $G$. This establishes the minimum admissible interaction radius, ensuring that every decision variable participates in the reduced optimization
problem. Preserving vertex participation alone,
however, does not guarantee that the resulting QAOA circuit can be executed within the available hardware constraints. We therefore impose an additional condition
on the transpiled two-qubit circuit depth.

\begin{figure}
    \centering
    \includegraphics[width=\linewidth]{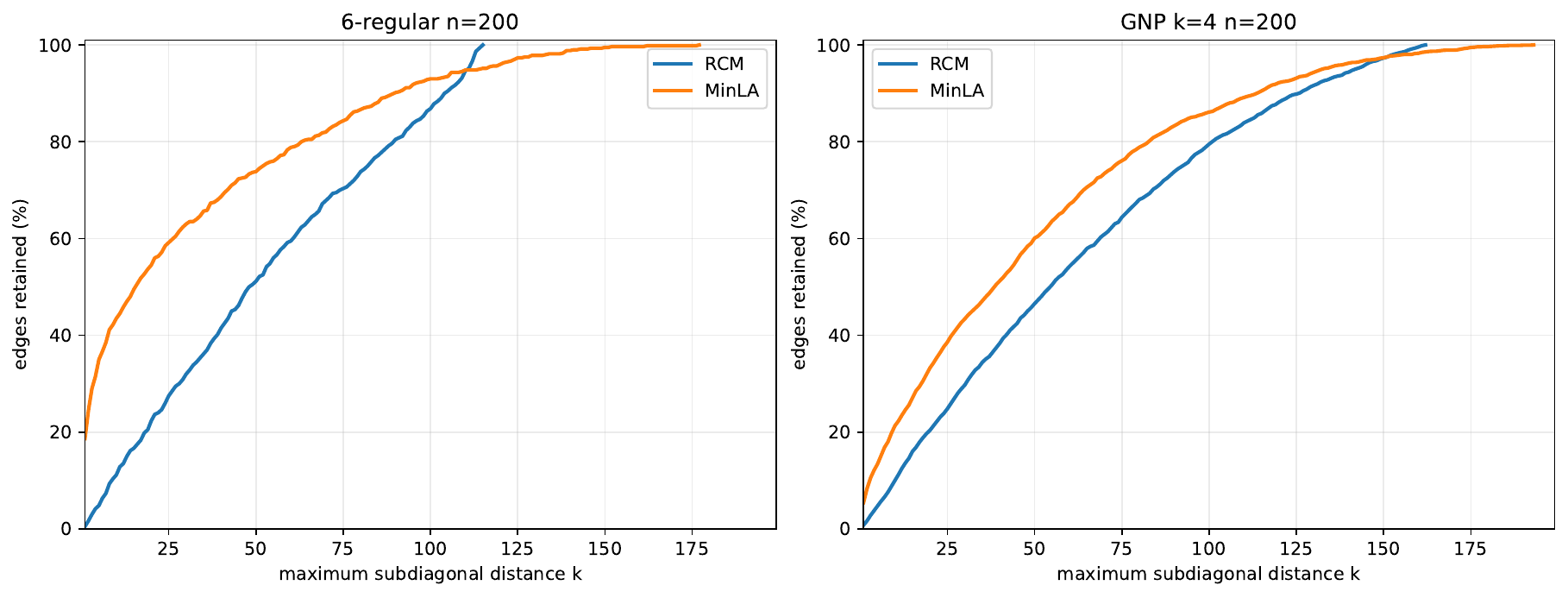}
    \caption{Cumulative edge retention $F(k)=|E_k|/|E|$ as a function of the interaction radius $k$ for MinLA and RCM vertex orderings of (a) a 6-regular graph and (b) a $G(n,p)$ graph.}
    \label{fig:edge-retention}
\end{figure}

Let $D_{\mathrm{2q}}(p,k)$ denote the two-qubit depth of a $p$-layer QAOA circuit constructed using $G_k$ and transpiled for the target hardware under a fixed compilation configuration. Let $b(\pi_{\mathrm{LA}}) := \max_{(u,v)\in E} |\pi_{\mathrm{LA}}(u)-\pi_{\mathrm{LA}}(v)|$ denote the maximum edge span under the chosen ordering. Then, given a prescribed depth cutoff $L$, we select the radius
    \begin{equation} \label{eq:optimal-interaction-radius}
        k^\star(p,L)=
        \max_{{
        k_{\mathrm{cov}}\leq k\leq b(\pi_{\mathrm{LA}})\\
        D_{\mathrm{2q}}(p,k)<L
        }} k,
    \end{equation}
The integer $k^\star(p,L)$ depends on the QAOA
depth $p$ as well as the hardware depth limit
$L$. Increasing $p$ generally increases the
circuit depth, potentially requiring a
smaller interaction radius to satisfy the
same budget. Moreover, although the retained
edge sets are nested, the transpiled depth
need not vary monotonically with $k$ due
to compilation heuristics. We therefore
select the largest feasible radius among
the evaluated candidates, rather than
assuming a monotonic relationship between
interaction count and compiled depth. For each candidate radius, the corresponding QAOA circuit is constructed and evaluated using the hardware-aware compilation pipeline described in Section~\ref{subsec:circuit-compilation}. The resulting \textit{hardware-feasible surrogate Hamiltonian} is given by
    \begin{equation} \label{eq:surrogate-hamiltonian}
        H_C^{(p,L)}
        =\frac{1}{2}
        \sum_{(u,v)\in E_{k^\star(p,L)}}
        (I-Z_uZ_v).
    \end{equation}
The fraction of interactions retained from the
original graph is determined
jointly by the graph structure, QAOA depth, and hardware
depth budget, rather than by a prescribed sparsification
ratio,
    \begin{equation} \label{eq:interaction-retention}
        \eta(p,L)=
        \frac{|E_{k^\star(p,L)}|}{|E|}.
    \end{equation}
To examine the influence of vertex ordering on this edge
retention fraction, Fig.~\ref{fig:edge-retention} compares the cumulative retention fraction
$F(k)=|E_k|/|E|$ for MinLA and RCM orderings for the representative graphs in Fig.~\ref{fig:minla-filtering}. We observe that MinLA exhibits substantially greater retention at smaller interaction radii. At $k=10$, MinLA retains $43.5\%$ of the edges compared with $11.2\%$ for RCM on the 6-regular graph, and $21.3\%$ compared with
$10.0\%$ on the $G(n,p)$ graph. Moreover, MinLA reaches
the minimum radius required to avoid isolated vertices
at smaller values of $k$ in both instances. These
observations support the use of MinLA for concentrating
interactions within narrow diagonal bands. Equivalently, for any target edge-retention ratio, MinLA often achieves that ratio with a smaller interaction radius than RCM. This is advantageous for hardware-constrained QAOA because a smaller radius generally leads to lower routing overhead and hence lower circuit depth, while preserving the same fraction of the original graph structure. However,
minimizing the cumulative edge span does not guarantee
greater retention at every radius, and the retention
curves may intersect at larger values of $k$.

\subsection{From Surrogate Hamiltonians to Original-Graph Solutions}
\label{subsec:surrogate-solutions}

The surrogate Hamiltonian is not intended to replace the original optimization problem. Rather, it provides a hardware-feasible approximation that retains a substantial fraction of the original graph structure while satisfying the depth constraints of the target quantum processor. As shown in the previous subsections, the combination of MinLA reordering and band-limited interaction selection preserves significantly more interactions than aggressively reducing the graph. Consequently, solutions sampled from the surrogate Hamiltonian can still capture important structural information about the original MAX-CUT instance.

However, the surrogate Hamiltonian does not contain every edge of the original graph. The omitted interactions do not participate in the quantum evolution and therefore do not influence the objective optimized during circuit execution. As a result, a bitstring that performs well on the surrogate Hamiltonian need not be optimal with respect to the full graph, since some interactions were never considered during quantum execution. The discrepancy arises not because the surrogate represents a different optimization problem, but because it only captures part of the original objective.

Let $E_s$ denote the set of retained interactions and $E_r = E \setminus E_s$ denote the omitted interactions. The original \textsc{MAX-CUT} objective can then be decomposed as

\begin{equation}
C(\mathbf x)
=
\sum_{(u,v)\in E_s} w_{uv}(x_u \oplus x_v)
+
\sum_{(u,v)\in E_r} w_{uv}(x_u \oplus x_v),
\label{eq:cost-decomposition}
\end{equation}

where the first term is represented in the surrogate Hamiltonian and influences the quantum state preparation, while the second term corresponds to interactions that were excluded to satisfy the hardware depth constraint. Note that for this study we take $w_{uv} = 1$ $\forall$ $u, v$.

The role of the quantum stage is therefore to identify promising regions of the solution space using an executable approximation of the original problem. The omitted interactions are subsequently reintroduced during a classical recovery phase, which evaluates and refines candidate solutions against the complete graph. In this way, the surrogate Hamiltonian provides a hardware-feasible starting point for optimization, while recovery accounts for the information that was incorporated in the hardware tailored quantum circuit or was corrupted from noise during quantum execution.

The next section develops recovery procedures that exploit the candidate bitstrings generated by the surrogate Hamiltonian and improve them with respect to the full \textsc{MAX-CUT} objective.

\section{Configuration Recovery for MAX-CUT}
\label{sec:recovery}

The MAX-CUT objective is a sum of edge contributions, making changes in the status of incident edges a natural basis for refining sampled configurations. We use this structure to develop two classical recovery procedures that explore different neighborhoods of a configuration, accepting an update only when it increases the full-graph objective $C_G$.

We rank the distinct QPU-sampled configurations by their values of $C_G$ and select the $B$ highest-scoring configurations as recovery seeds. Let $\mathcal X_B$ denote this set, or all distinct samples if fewer than $B$ are available. For a recovery procedure $\mathcal R$, the returned assignment is
    \begin{equation}
        \mathbf x_{\mathrm{rec}}
        \in
        \operatorname*{arg\,max}_{\mathbf y\in
        \{\mathcal R(\mathbf x):\,\mathbf x\in\mathcal X_B\}}
        C_G(\mathbf y).
        \label{eq:recovered-assignment}
    \end{equation}
Because each trajectory retains its initial assignment unless an improvement is found,
    \begin{equation}
        C_G(\mathbf x_{\mathrm{rec}})
        \geq
        \max_{\mathbf x\in\mathcal X_B}C_G(\mathbf x).
        \label{eq:recovery-monotonicity}
    \end{equation}
The trajectories are independent and can run in parallel on classical processors without further QPU calls.

\subsection{Signed Vertex Residues}
\label{subsec:signed-residue}

Flipping a vertex changes the cut status of every incident edge: previously uncut edges become cut, while previously cut edges become uncut. An improving move must therefore balance both contributions. For a vertex $i\in V$, we define
    \begin{align}
        B_i(\mathbf x)
        &=
        \sum_{j\in N_G(i)}\mathbf 1[x_i=x_j],\\
        G_i(\mathbf x)
        &=
        \sum_{j\in N_G(i)}\mathbf 1[x_i\neq x_j],
    \end{align}
where $B_i$ and $G_i$ count its incident uncut (\textit{bad}) and cut (\textit{good}) edges, respectively. We note that the neighborhood $N_G(i)$ is taken in the original graph, so the gain accounts for both retained and omitted interactions. The \textit{signed vertex residue} is
    \begin{equation}
        \rho_i(\mathbf x)
        =
        B_i(\mathbf x)-G_i(\mathbf x).
        \label{eq:vertex-residue}
    \end{equation}
Since only edges incident on $i$ change their cut status, this residue is the gain obtained from flipping $i$,
    \begin{equation}
        \rho_i(\mathbf x) = C_G(\mathbf x\oplus\mathbf e_i)-C_G(\mathbf x),
        \label{eq:vertex-gain}
    \end{equation}
where $\mathbf e_i$ is the binary unit vector at $i$ and $\oplus$ denotes component-wise addition $\mod 2$. Thus, $\rho_i>0$ identifies a strictly improving single-vertex move.  This identity applies to a single flip at the current assignment; gains must be updated after a move and cannot generally be added to assess simultaneous flips of adjacent vertices.

\subsection{Greedy Signed-Residue Recovery}
\label{subsec:greedy-recovery}

We first propose a greedy recovery using signed vertex residues to perform a steep ascent on the full-graph cut. At every iteration, vertices are ranked by the residues calculated from the QPU-sampled configuration, and the highest-ranked vertex is flipped if its residue is positive. After an accepted flip, the residues must be updated to reflect the new configuration before selecting the next move. Since flipping vertex $i$ changes only its incident edges, only the residues of $i$ and its neighbors are affected. The following lemma gives an incremental update that avoids recomputing residues over the full graph. The proof is outlined in Appendix \ref{sec:proof-of-lem-res-update}.

\begin{lemma}[Local residue update]
    \label{lem:residue-update}
    Let $\mathbf x\in\{0,1\}^{|V|}$ be the QPU-sampled configuration, and $\mathbf x'=\mathbf x\oplus\mathbf e_i$ be the configuration obtained by flipping the assignment of vertex $i$. The signed vertex residues satisfy
        \begin{equation}
            \rho_j(\mathbf x')=
            \begin{cases}
                -\rho_i(\mathbf x), & j=i,\\
                \rho_j(\mathbf x)-2\bigl(1-2\mathbf 1[x_i\neq x_j]\bigr),
                & j\in N_G(i),\\
                \rho_j(\mathbf x), & \text{otherwise}.
            \end{cases}
        \end{equation}
\end{lemma}

\begin{algorithm}[ht]
    \caption{Greedy signed-residue recovery}
    \label{alg:greedy-recovery}
    \begin{algorithmic}[1]
        \Require Original graph $G=(V,E)$, initial configuration
        $\mathbf x^{(0)}$, flip budget $F_{\max}$
        \Ensure Refined configuration $\mathbf x$
        \State $\mathbf x\gets\mathbf x^{(0)}$
        \State Compute $\rho_i\gets B_i(\mathbf x)-G_i(\mathbf x)$
        for all $i\in V$
        \For{$f=1,\ldots,F_{\max}$}
            \State $i^\star\gets
            \operatorname*{arg\,max}_{i\in V}\rho_i$
            \Comment{Break ties by a fixed vertex order}
            \If{$\rho_{i^\star}\leq0$}
                \State \textbf{break}
            \EndIf
            \For{$j\in N_G(i^\star)$}
                \State $\rho_j\gets\rho_j-
                2\bigl(1-2\mathbf 1[x_{i^\star}\neq x_j]\bigr)$
            \EndFor
            \State $\rho_{i^\star}\gets-\rho_{i^\star}$
            \State $x_{i^\star}\gets1-x_{i^\star}$
        \EndFor
        \State \Return $\mathbf x$
    \end{algorithmic}
\end{algorithm}

The algorithm terminates when no improving single-vertex move remains or when $F_{\max}$ flips have been accepted. This is outlined in Algorithm \ref{alg:greedy-recovery} for one seed; recovery trajectories are run independently for all $\mathbf x^{(0)}\in\mathcal X_B$. The relation between signed residue and cut gain yields two guarantees: accepted moves improve the objective, and the bounded, integer-valued objective limits how many such moves can occur.

\begin{theorem}[Monotonicity]
\label{thm:greedy-monotonicity}
Suppose Algorithm~\ref{alg:greedy-recovery} accepts $F$ flips, and let $\mathbf x^{(t)}$ denote the configuration after $t$ accepted flips, with $\mathbf x^{(0)}$ the initial configuration. Then
\begin{equation}
    C_G(\mathbf x^{(t+1)})
    >
    C_G(\mathbf x^{(t)}),
    \qquad t=0,\ldots,F-1.
\end{equation}
Consequently, the returned configuration satisfies
\begin{equation}
    C_G(\mathbf x^{(F)})
    \geq C_G(\mathbf x^{(0)}),
\end{equation}
with equality if and only if $F=0$, regardless of whether recovery terminates because no improving flip remains or because the flip budget is exhausted.
\end{theorem}

The proof is outlined in Appendix \ref{sec:proof-of-thm-greedy-mono}. The strict improvement of the cut objective ensures
that greedy recovery cannot continue indefinitely.
The following result characterizes its termination
and the condition under which the returned
configuration is locally optimal. The proof is detailed in Appendix \ref{sec:proof-of-thm-greedy-term}.

\begin{theorem}[Finite termination and local optimality]
\label{thm:greedy-termination}
Starting from an initial configuration $\mathbf{x}^{(0)}$, Algorithm
~\ref{alg:greedy-recovery} accepts at most
    \begin{equation} \label{eq:flip-bound}
    F\leq\min\left\{
    F_{\max},\,
    |E|-C_G(\mathbf{x}^{(0)})
    \right\}
    \end{equation}
flips. Furthermore, if the returned configuration
$\widehat{\mathbf{x}}$ satisfies
$\max_{i\in V}\rho_i(\widehat{\mathbf{x}})\leq0$,
then it is a one-flip local optimum of the
original MAX-CUT objective.
\end{theorem}

\begin{corollary}[Cut guarantee at local optimality]
\label{cor:one-flip-guarantee}
If $\widehat{\mathbf{x}}$ is a one-flip local
optimum, then
\begin{equation}
C_G(\widehat{\mathbf{x}})
\geq\frac{|E|}{2}
\geq\frac{C_G^\star}{2},
\label{eq:local-optimum-bound}
\end{equation}
where $C_G^\star$ denotes the optimal
MAX-CUT value.
\end{corollary}

The proof is presented in Appendix \ref{sec:proof-of-cor-sing-flip}. The above guarantee holds independently of the
initial configuration and follows solely from
the classical local-optimality condition.
If recovery terminates upon exhausting
$F_{\max}$, monotonic improvement is still
guaranteed, but local optimality cannot
be asserted without checking the final
residues.

\subsubsection{Limitations of Single-Vertex Recovery}
\label{subsec:single-flip-limitation}

Although greedy signed-residue recovery guarantees
monotonic improvement, restricting updates to
individual vertices can cause the algorithm to
terminate at a suboptimal configuration. In
particular, the absence of an improving single-vertex
flip does not preclude an improvement through
simultaneous changes to multiple assignments.

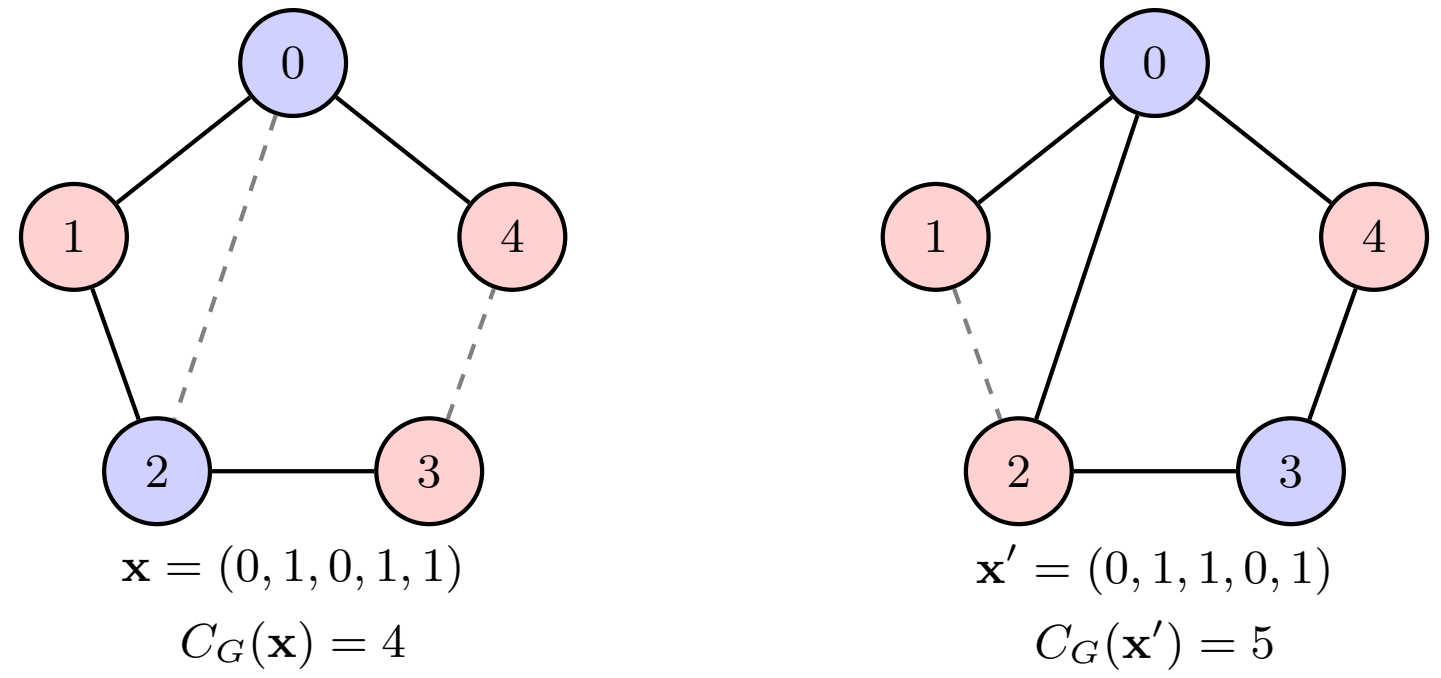
\begin{figure}[htbp]
\centering
\begin{tikzpicture}[
    scale=1,
    vertex/.style={
        circle, draw=black, thick,
        minimum size=7mm, inner sep=0pt,
        font=\small
    },
    partA/.style={vertex,fill=blue!18},
    partB/.style={vertex,fill=red!18},
    cut/.style={black,thick},
    uncut/.style={gray,dashed,thick}
]

\begin{scope}[shift={(0,0)}]
    \node[partA] (a0) at (0,1.7) {$0$};
    \node[partB] (a1) at (-1.45,0.55) {$1$};
    \node[partA] (a2) at (-0.9,-1.0) {$2$};
    \node[partB] (a3) at (0.9,-1.0) {$3$};
    \node[partB] (a4) at (1.45,0.55) {$4$};

    \draw[cut] (a0)--(a1);
    \draw[cut] (a1)--(a2);
    \draw[cut] (a2)--(a3);
    \draw[uncut] (a3)--(a4);
    \draw[cut] (a4)--(a0);
    \draw[uncut] (a0)--(a2);

    \node[draw=none,font=\small] at (0,-1.65)
        {$\mathbf{x}=(0,1,0,1,1)$};
    \node[draw=none,font=\small] at (0,-2.15)
        {$C_G(\mathbf{x})=4$};
\end{scope}

\begin{scope}[shift={(5.7,0)}]
    \node[partA] (b0) at (0,1.7) {$0$};
    \node[partB] (b1) at (-1.45,0.55) {$1$};
    \node[partB] (b2) at (-0.9,-1.0) {$2$};
    \node[partA] (b3) at (0.9,-1.0) {$3$};
    \node[partB] (b4) at (1.45,0.55) {$4$};

    \draw[cut] (b0)--(b1);
    \draw[uncut] (b1)--(b2);
    \draw[cut] (b2)--(b3);
    \draw[cut] (b3)--(b4);
    \draw[cut] (b4)--(b0);
    \draw[cut] (b0)--(b2);

    \node[draw=none,font=\small] at (0,-1.65)
        {$\mathbf{x}'=(0,1,1,0,1)$};
    \node[draw=none,font=\small] at (0,-2.15)
        {$C_G(\mathbf{x}')=5$};
\end{scope}

\end{tikzpicture}

\caption{(L) The initial configuration is a one-flip
local optimum with cut value $4$.
(R) Simultaneously flipping vertices $2$
and $3$ produces the globally optimal cut
of value $5$. Blue and red vertices indicate
the two partitions, while solid and dashed
edges represent cut and uncut edges,
respectively.}
\label{fig:single-flip-example}
\end{figure}

For example, consider the five-vertex graph illustrated in
Fig.~\ref{fig:single-flip-example}. For the initial configuration $\mathbf{x}=(0,1,0,1,1)$, the cut value is
$C_G(\mathbf{x})=4$, with signed residues $\boldsymbol{\rho}(\mathbf{x})
=(-1,-2,-1,0,0).$

\begin{table}[htbp]
\centering
\caption{Effect of single-vertex flips on the
initial configuration
$\mathbf{x}=(0,1,0,1,1)$, with
$C_G(\mathbf{x})=4$.}
\label{tab:single-flip-example}
\small
\begin{tabular}{ccc}
\toprule
Vertex $i$ &
Gain $\rho_i$ &
$C_G(\mathbf{x}\oplus\mathbf e_i)$\\
\midrule
0 & $-1$ & 3\\
1 & $-2$ & 2\\
2 & $-1$ & 3\\
3 & $0$  & 4\\
4 & $0$  & 4\\
\bottomrule
\end{tabular}
\end{table}

As shown in Table~\ref{tab:single-flip-example},
every possible single-vertex flip yields a
nonpositive gain. Consequently, strict greedy
recovery terminates at this one-flip local optimum.
However, simultaneously flipping vertices $2$
and $3$ produces the configuration
$\mathbf{x}'=(0,1,1,0,1)$, increasing the cut
value to $C_G(\mathbf{x}')=5$. This is globally
optimal, since the graph contains a triangle
and hence at least one of its six edges must
remain uncut. 

This example illustrates that strict single-vertex ascent can become trapped at a local optimum,
requiring coordinated flips wherein a subset of
vertices is optimized jointly to yield a better solution. This motivates the \textit{block-based recovery strategy} developed in the subsequent subsection.


\subsection{Bipartite-Block Recovery via Minimum Cuts}
\label{subsec:block-recovery}

Jointly optimising an arbitrary block (subset) of vertices while keeping the remaining assignments of the remaining vertices fixed itself can be computationally
difficult, as the resulting subproblem may have the
same complexity as \textsc{MAX-CUT}. To overcome this limitation, we construct bipartite blocks for which the conditional optimization can be transformed into an
$s$--$t$ minimum-cut problem ~\citep{kolmogorov2004energy}. We derive the reduction step by step below.

\subsubsection{Construction of Bipartite Blocks.}
First, we do a greedy vertex coloring of the original graph $G$, yielding $q$ disjoint color classes
    \begin{equation} \label{eq:color-partition}
    V=C_1\mathbin{\dot\cup}\cdots
    \mathbin{\dot\cup}C_q,
    \end{equation}
where vertices belonging to the same color class are non-adjacent. For each pair of distinct color classes, we define
    \begin{equation}
    S_{ab} := C_a\cup C_b,
    \qquad 1\leq a<b\leq q.
    \label{eq:bipartite-block}
    \end{equation}
This coloring remains fixed throughout recovery,
providing $Q=\binom{q}{2}$ candidate blocks that can be reused across all QPU-sampled configurations. Since every edge joining vertices in $S_{ab}$ must connect the two distinct color classes, the induced graph $G[S_{ab}]$ is bipartite. This bipartite structure, as we show subsequently, provides a useful basis for simplifying the conditional optimization.

\subsubsection{Conditional optimization.}

We consider a bipartite block $S:=S_{ab}=C_a\cup C_b$ and its complement $\bar S=V\setminus S$. Given the current configuration $\mathbf{x}$, we seek an assignment $\mathbf{y}_S$ that maximizes the cut value of the original graph while keeping the assignments outside $S$ fixed, i.e., 
    \begin{equation} \label{eq:conditional-block-maxcut}
        \mathbf{y}_S^\star
        \in
        \operatorname*{arg\,max}_{
        \mathbf{y}_S\in\{0,1\}^{|S|}
        }
        C_G(\mathbf{y}_S,\mathbf{x}_{\bar S}).
    \end{equation}
To express the above objective in terms of optimisation variables, we partition the edge set into
three categories: edges internal to $S$, edges
internal to $\bar S$, and edges connecting the
two. Since the exterior assignments remain fixed,
edges entirely within $\bar S$ contribute a
constant and can be omitted from the optimization.
The resulting conditional objective is
\begin{equation} \label{eq:conditional-objective}
    F_S(\mathbf{y}_S;\mathbf{x}_{\bar S})
    = 
    \sum_{(u,v)\in E[S]}
    \mathbf{1}[y_u\neq y_v]
    +
    \sum_{u\in S}
    \sum_{v\in N_G(u)\cap\bar S}
    \mathbf{1}[y_u\neq x_v].
\end{equation}

The first term in \eqref{eq:conditional-objective} captures the cut contribution of edges within the block, whereas the second term accounts for edges connecting the block to its fixed exterior. Both contributions must be
considered jointly, since improving the internal cut may come at the expense of boundary edges. We next show how each term can be transformed so that their joint optimization reduces to a tractable form.

\paragraph{Step 1: Transforming internal interactions:}
We consider the first term in ~\eqref{eq:conditional-objective}. Since $G[S]$ is bipartite, its vertices can
be assigned reference labels according to
their color classes:
    \begin{equation} \label{eq:bipartition-labels}
    c_u=
    \begin{cases}
    0, & u\in C_a,\\
    1, & u\in C_b.
    \end{cases}
    \end{equation} 
We do a transformation of variable,
    \begin{equation} \label{eq:variable-transformation}
        t_u=y_u\oplus c_u,
        \qquad u\in S,
    \end{equation}
where $\oplus$ denotes binary XOR. This transformation complements the assignments on one side of the bipartition. To see what this variable is meant to do, consider an internal edge $(u,v)\in E[S]$. Since its endpoints belong to different color classes, we have that $c_u \neq c_v$, i.e., $c_u\oplus c_v=1$. Therefore,
    \begin{equation}
        \begin{aligned}
        y_u\oplus y_v
        =(t_u\oplus c_u)
          \oplus(t_v\oplus c_v) =t_u\oplus t_v\oplus1.
        \end{aligned}
    \end{equation}
Cutting an internal edge $(u,v)\in E[S]$ becomes agreeing across that edge in the transformed variables,
\begin{equation}
\mathbf{1}[y_u\neq y_v]
=
1 - \mathbf{1}[t_u \neq t_v].
\label{eq:internal-edge-transform}
\end{equation}
Summing over all internal edges yields,
\begin{equation}
\sum_{(u,v)\in E[S]}
\mathbf{1}[y_u\neq y_v]
=
|E[S]|-
\sum_{(u,v)\in E[S]}
\mathbf{1}[t_u\neq t_v].
\label{eq:internal-transformed}
\end{equation}
Thus, after dropping the constant $|E[S]|$, maximizing the internal cut is equivalent to minimizing disagreements
between the transformed variables $\sum_{(u,v)\in E[S]}
\mathbf{1}[t_u\neq t_v]$. In particular, each internal edge now contributes a non-negative penalty whenever its endpoints receive different transformed labels.

\paragraph{Step 2: Transforming boundary interactions:}

We consider the second term in ~\eqref{eq:conditional-objective}. Unlike internal edges, each boundary edge
connects a variable vertex $u\in S$ to
a vertex $v\in\bar S$ whose assignment
$x_v$ is fixed. Consequently, its
contribution depends only on the chosen
assignment of $u$. For each $u\in S$, we define the boundary reward in terms of the transformed variable
$t_u$. Since assigning $t_u=r$ corresponds
to the original assignment $y_u=r\oplus c_u$,
the number of boundary edges cut is
    \begin{equation} \label{eq:boundary-reward}
    H_u(r)=
    \sum_{v\in N_G(u)\cap\bar S}
    \mathbf{1}[r\oplus c_u\neq x_v],
    \qquad r\in\{0,1\}.
    \end{equation}
Here, $H_u(r)$ counts cut weight contributed by edges from $u$ to the fixed outside vertices. Since our objective is to maximize the boundary reward, we convert it into an equivalent non-negative minimization cost. For each vertex, we define,
    \begin{align} \label{eq:boundary-penalty}
    K_u := \max\{&H_u(0),H_u(1)\},\\
    D_u(r) := &K_u-H_u(r).
    \end{align}
The unary penalties $D_u(r)\geq0$ represent the loss in boundary reward
relative to the maximum reward attainable
by vertex $u$ individually. Hence,
\begin{equation}
    \sum_{u\in S}H_u(t_u)
    =
    \sum_{u\in S}K_u
    -
    \sum_{u\in S}D_u(t_u).
\end{equation}
Combining the transformed internal interactions
in Eq.~\eqref{eq:internal-transformed} with
the boundary penalties, the conditional
MAX-CUT objective can be expressed as
\begin{equation}
F_S(\mathbf{y}_S;\mathbf{x}_{\bar S})
=
|E[S]|+\sum_{u\in S}K_u
-\mathcal E_S(\mathbf t),
\end{equation}
where
\begin{equation}
\mathcal E_S(\mathbf t)=
\sum_{(u,v)\in E[S]}
\mathbf{1}[t_u\neq t_v]
+\sum_{u\in S}D_u(t_u).
\label{eq:block-mincut-energy}
\end{equation}
Since the remaining terms are independent of $\mathbf t$, maximizing the conditional cut is equivalent to minimizing
$\mathcal E_S(\mathbf t)$.

\paragraph{Step 3: Reduction to an $s-t$ minimum cut:}

In this step, we show how the minimization in \eqref{eq:block-mincut-energy} is solved. The pairwise penalties take the form $V(t_u,t_v)=\mathbf{1}[t_u\neq t_v]$ and satisfy the binary submodularity condition, i.e., the combined cost of the two equal-label assignments must not exceed the combined cost of the two unequal-label assignments.
    \begin{equation}
    V(0,0)+V(1,1)
    \leq V(0,1)+V(1,0),
    \end{equation}
Together with the unary penalties, the resulting energy admits an exact $s-t$ minimum-cut representation \citep{kolmogorov2004energy}. We construct a flow network containing one vertex for each $u\in S$, together with a source $s$ and a sink $t$. For each vertex $u\in S$, we introduce two directed edges with capacities,
    \begin{equation} \label{eq:terminal-capacities}
        \operatorname{cap}(s,u)=D_u(1),
        \qquad
        \operatorname{cap}(u,t)=D_u(0).
    \end{equation}
For every internal edge $(u,v)\in E[S]$, we introduce two oppositely directed edges, $(u,v)$ and $(v,u)$, each with unit capacity. An $s-t$ cut partitions the vertices of the flow network into two sets, one containing the source $s$ and the other containing the sink $t$. We associate this partition with the transformed binary variables as
    \begin{equation} \label{eq:mincut-labels}
        t_u=
        \begin{cases}
        0, & u\text{ belongs to the source side},\\
        1, & u\text{ belongs to the sink side}.
        \end{cases}
    \end{equation}
Under this convention, assigning $t_u=0$ places $u$ on the source side, so the edge $(u,t)$ crosses the cut and contributes $D_u(0)$. Similarly, assigning $t_u=1$ places $u$ on the sink side, so the edge $(s,u)$ contributes $D_u(1)$.

For an internal edge $(u,v)\in E[S]$, the corresponding vertices lie on opposite sides of the cut precisely when $t_u\neq t_v$. In this case, exactly one of the two directed edges contributes unit capacity; otherwise, neither contributes. Thus, the total capacity of a cut corresponding to $\mathbf t$ is
    \begin{equation} \label{eq:mincut-capacity}
        \operatorname{CutCap}(\mathbf t)
        =\sum_{u\in S}D_u(t_u)
        +\sum_{(u,v)\in E[S]}
        \mathbf{1}[t_u\neq t_v]
        = \mathcal E_S(\mathbf t).
    \end{equation}
Since the cut capacity is exactly the transformed energy, finding a minimum $s-t$ cut yields an optimal solution to \eqref{eq:block-mincut-energy}. To obtain this solution, we first compute a maximum $s-t$ flow and then identify the set $U$ of block vertices reachable from $s$ through edges with positive residual capacity. These vertices lie on the source side of a minimum cut, while the remaining vertices lie on the sink side. The optimal transformed assignment is therefore,
    \begin{equation}  \label{eq:mincut-optimal-labels}
        t_u^\star=
        \begin{cases}
            0, & u\in U,\\
            1, & u\in S\setminus U.
        \end{cases}
    \end{equation}
Finally, reversing the transformation in
\eqref{eq:variable-transformation}, we obtain
the corresponding block assignments
\begin{equation}
y_u^\star=t_u^\star\oplus c_u,
\qquad u\in S.
\label{eq:recover-block-labels}
\end{equation}
The resulting $\mathbf y_S^\star$ maximizes
the original MAX-CUT objective over the
selected block, with all assignments
outside $S$ held fixed.

\begin{algorithm}[t]
\caption{Bipartite-block configuration recovery}
\label{alg:block-recovery}
\begin{algorithmic}[1]
\Require Graph $G=(V,E)$,
color classes $\{C_a\}_{a=1}^q$,
initial configuration $\mathbf{x}^{(0)}$
\Ensure Recovered configuration $\mathbf{x}$

\State $\mathbf{x}\gets\mathbf{x}^{(0)}$

\Repeat
    \State $\mathrm{improved}\gets\mathrm{false}$

    \For{each block $S=C_a\cup C_b$, $a<b$}
        \State Construct the
        flow network for $S$
        \State Compute a maximum $s$--$t$ flow
        \State Find the source-reachable set
        $ U$ in the network
        \State Set
        $t_u^\star=0$ if $u\in U$,
        and $t_u^\star=1$ otherwise
        \State Recover
        $y_u^\star=t_u^\star\oplus c_u$
        for each $u\in S$

        \If{$C_G(\mathbf{y}_S^\star,
            \mathbf{x}_{\bar S})>C_G(\mathbf{x})$}
            \State $\mathbf{x}_S
            \gets\mathbf{y}_S^\star$
            \State $\mathrm{improved}
            \gets\mathrm{true}$
            \State \textbf{break}
        \EndIf
    \EndFor

\Until{$\mathrm{improved}=\mathrm{false}$}

\State \Return $\mathbf{x}$
\end{algorithmic}
\end{algorithm}

\subsubsection{Recovery Procedure and Guarantees}

Bipartite-block recovery is applied independently to each selected QPU-sampled configuration, without requiring a preceding single-vertex greedy recovery stage (Algorithm~\ref{alg:greedy-recovery}). Starting from a configuration $\mathbf{x}$, we rank the candidate blocks according to their aggregate signed vertex residues,
    \begin{equation} \label{eq:block-ranking}
        R_{ab}(\mathbf{x})
        =
        \sum_{u\in S_{ab}}\rho_u(\mathbf{x}),
    \end{equation}
where $\rho_u(\mathbf{x})$ is the gain obtained by flipping vertex $u$ individually. Blocks with larger residues are examined first, as they indicate a greater aggregate tendency toward improving single-vertex assignments. 

For each block $S$, we solve the conditional optimization exactly using the $s-t$ minimum-cut construction described previously. The improvement in the original objective is
    \begin{equation} \label{eq:block-gain}
        \Delta_S^\star(\mathbf{x})
        =
        C_G(\mathbf{y}_S^\star,\mathbf{x}_{\bar S})
        -C_G(\mathbf{x}).
    \end{equation}
Since the current block assignment is feasible for the conditional optimization, $\Delta_S^\star(\mathbf{x})\geq0$. We accept an update only when $\Delta_S^\star(\mathbf{x})>0$. After each accepted update, the block residues are recomputed and the blocks are ranked again. The procedure terminates when a complete pass over all candidate blocks produces no improvement.

Every accepted update strictly increases the
integer-valued cut objective. Since
$C_G(\mathbf{x})\leq |E|$, the number of
accepted updates is bounded by
\begin{equation}
A\leq |E|-C_G(\mathbf{x}^{(0)}).
\label{eq:block-termination}
\end{equation}
Thus, recovery terminates in finitely many
updates without requiring an explicit budget.
Moreover, with $Q=\binom{q}{2}$ candidate
blocks, at most $Q(A+1)$ conditional block
optimizations are attempted. Since each
minimum-cut problem is solvable in polynomial
time, the overall recovery procedure also
has polynomial worst-case complexity.

The final configuration is locally optimal
with respect to every candidate block.
In particular, it is also locally optimal
under single-vertex flips, since any improving
single-vertex assignment would be feasible
within a block containing that vertex.
Consequently, for unweighted \textsc{MAX-CUT},
the recovered configuration satisfies
\begin{equation}
C_G(\mathbf{x}_{\mathrm{final}})
\geq \frac{|E|}{2}
\geq \frac{C_G^\star}{2}.
\label{eq:block-approx-guarantee}
\end{equation}
Thus, while the procedure does not guarantee
a globally optimal cut, it provides a
$1/2$-approximation guarantee upon convergence,
in addition to monotonic improvement from
the initial configuration. Finally, recovery is performed independently
for each selected QPU-sampled configuration,
and the best recovered solution is retained.
This allows direct comparison with single-vertex greedy recovery using identical initial
configurations, without additional QPU
executions.

\section{Experimental Evaluation on \texttt{ibm\_kingston}}
\label{subsec:experimental-protocol}

In this section, we evaluate the effectiveness of the proposed hybrid quantum--classical workflow for non-hardware-native \textsc{MAX-CUT} instances. Specifically, we investigate
three aspects:
\begin{enumerate}
    \item \textit{Circuit-depth reduction:} The extent to which surrogate Hamiltonians reduce
    transpiled two-qubit depth while satisfying hardware-specific depth constraints.
    \item \textit{Solution quality from surrogate sampling:}
    The ability of reduced QAOA circuits to generate configurations with high cut values on the original graph.
    \item \textit{Effectiveness of classical recovery:}
    The improvement in solution quality achieved by refining candidate configurations against the original \textsc{MAX-CUT} objective.
\end{enumerate}

We emphasise that our primary focus is not comparison against state-of-the-art classical optimization techniques, but to assess whether hardware-constrained surrogate execution,
combined with classical recovery, enables effective QAOA-based optimization of graph instances whose full cost Hamiltonians exceed practical circuit depth budgets.

\paragraph{Experimental Setup:}

We evaluate these aspects on 120 benchmark instances from HAMLIB~\cite{sawaya2024hamlib}, comprising 20 instances each of 6-regular and Erd\H{o}s--R'enyi ($G(n,p)$) graphs at each problem size $n\in{40,60,100}$. For each instance, we execute both the original QAOA circuit and its corresponding MinLA-band surrogate on \texttt{ibm\_kingston}. The experimental configuration is summarized in
Table~\ref{tab:exp-setup}.

\begin{table}[t]
\centering
\caption{Experimental configuration}
\label{tab:exp-setup}
\small
\begin{tabularx}{\columnwidth}{@{}lX@{}}
\toprule
\textbf{Setting} & \textbf{Configuration} \\
\midrule
Quantum processor
    & \texttt{ibm\_kingston} \\
Graph families
    & 6-regular and $G(n,p)$ (HAMLIB) \\
Problem sizes
    & $n=40,60,100$ \\
QPU shots
    & 10,000 per circuit \\
Compilation
    & SATMapper~\cite{MATSUO_2023},
      commuting-gate routing~\cite{Weidenfeller2022};
      Qiskit optimization level 3 \\
Transpiler seed
    & 123 \\
Noise handling
    & Gate and measurement twirling enabled;
      dynamical decoupling disabled \\
QAOA layers
    & $p\in\{1,2,3\}$ \\
Depth budget $L$
    & 150 (6-regular); 200 ($G(n,p)$) \\
Recovery candidates
    & 50 distinct QPU-sampled bitstrings \\
\bottomrule
\end{tabularx}
\end{table}

\paragraph{Surrogate Selection and Depth Budgets:}
Vertex orderings are obtained using the weighted-aggregation MinLA heuristic~\cite{safro2006}. For each instance, we select the lexicographically largest feasible pair $(p,k)$, prioritizing the QAOA depth $p\in\{1,2,3\}$ before the interaction-band radius $k$. Feasibility requires that every vertex retains at least one incident edge and that the transpiled two-qubit depth does not exceed the prescribed budget $L$.

The family-specific depth budgets were chosen to accommodate the minimum-bandwidth surrogates that retain at least one incident edge per vertex across all tested instances. The maximum observed transpiled two-qubit depths of these surrogates, evaluated using a single QAOA layer, were 118 for 6-regular graphs and 190 for $G(n,p)$ graphs, motivating budgets of L=150 and L=200, respectively. This provides an empirical feasibility criterion for the tested graphs and compilation setup, rather than a universal depth guarantee.

\paragraph{Performance Metrics:}
All solution qualities are evaluated on the original graph $G$. Let $C_G(\boldsymbol{x})$
denote the cut value of configuration
$\boldsymbol{x}$ and $C_G^\star$ the optimal
cut value, certified using Gurobi for every
instance. We define the approximation ratio as
\begin{equation}
\label{eq:approx-ratio}
\operatorname{AR}(\boldsymbol{x})
=
\frac{C_G(\boldsymbol{x})}{C_G^\star}.
\end{equation}

\paragraph{Recovery Evaluation:}
To assess the improvement due to classical
recovery, we select 50 distinct starting
configurations $S_i^{50}$ from the surrogate
QAOA samples of instance $i$. Each configuration
is refined independently using recovery
method $R$. We compare the best cut values
before and after recovery using
\begin{align}
\operatorname{AR}_{i,\mathrm{raw}}^{(50)}
&=
\max_{\boldsymbol{x}\in S_i^{50}}
\operatorname{AR}(\boldsymbol{x}),
\\
\operatorname{AR}_{i,R}^{(50)}
&=
\max_{\boldsymbol{x}\in S_i^{50}}
\operatorname{AR}(R(\boldsymbol{x})).
\end{align}

This comparison isolates the improvement
obtained through classical recovery using
the same 50 starting configurations.
Results are reported as means and sample
standard deviations over the 20 instances
in each graph-family and size group.

\subsection{Results on 6-Regular Graphs}

Although regular graphs are sparse from a graph-theoretic perspective, they remain non-native to superconducting processors because their interaction structure differs substantially from the sparse near-linear coupling graph of the target hardware. Direct execution therefore incurs significant routing overhead, making these instances a useful testbed for evaluating whether hardware-aware surrogate Hamiltonians can preserve useful optimization information while remaining executable within realistic depth budgets.

\paragraph{Surrogate Construction and Circuit Depth:}

\begin{table}[htbp]
\centering
\footnotesize
\setlength{\tabcolsep}{4pt}
\renewcommand{\arraystretch}{1.2}
\begin{tabular}{|c|l|l|}
\hline
$p$ & $\boldsymbol{\beta}$ &
$\boldsymbol{\gamma}$ \\
\midrule
1 & $(0.39270)$
  & $(-0.42055)$ \\
2 & $(0.51825,\,0.27656)$
  & $(-0.32421,\,-0.57977)$ \\
3 & $(0.57265,\,0.39892,\,0.21694)$
  & $(-0.28142,\,-0.50403,\,-0.57681)$ \\
\hline
\end{tabular}
\caption{Fixed QAOA angles for 6-regular graphs}
\label{tab:regular6-angles}
\end{table}

Each 6-regular graph contains $|E|=3n$ interactions. We use the fixed-angle approach of \cite{fixedangle2021wurtz}, with QAOA parameters expressed in the positive-$ZZ$ convention of our implementation. The parameter values for $p=1,2,3$ are in Table \ref{tab:regular6-angles}.


Table~\ref{tab:regular6-minla},  shows that the selected MinLA bands retain a substantial fraction of these interactions, preserving $70.5\%$, $49.2\%$, and $51.4\%$ of the original edges for $n=40$, $60$, and $100$, respectively. The $p^\star$ column counts instances selected at each QAOA depth. Entries are means $\pm$ sample standard deviations; ranges are given in brackets.

\begin{table}[h]
\centering
\footnotesize
\setlength{\tabcolsep}{4pt}
\renewcommand{\arraystretch}{1.2}
\begin{tabular}{|r|c|c|c|c|}
\hline
$n$ & $\overline{C_G^\star}$ & $p^\star=1/2/3$ & $\overline{k^\star}$ [range] & Retained (\%) \\
\hline
40  & $92.3\pm1.2$  & $0/0/20$ & $8.5\pm1.1$ $[6,10]$  & $70.5\pm3.2$  \\
60  & $140.1\pm1.8$ & $0/2/18$ & $4.7\pm1.6$ $[3,10]$  & $49.2\pm5.8$  \\
100 & $237.4\pm1.9$ & $6/8/6$  & $10.2\pm8.6$ $[3,25]$ & $51.4\pm15.7$ \\
\hline
\end{tabular}
\caption{Selected MinLA bands for 20 $6$-regular instances per size.}
\label{tab:regular6-minla}
\end{table}

\begin{figure}[htbp]
    \centering
    \includegraphics[width=0.8\linewidth]{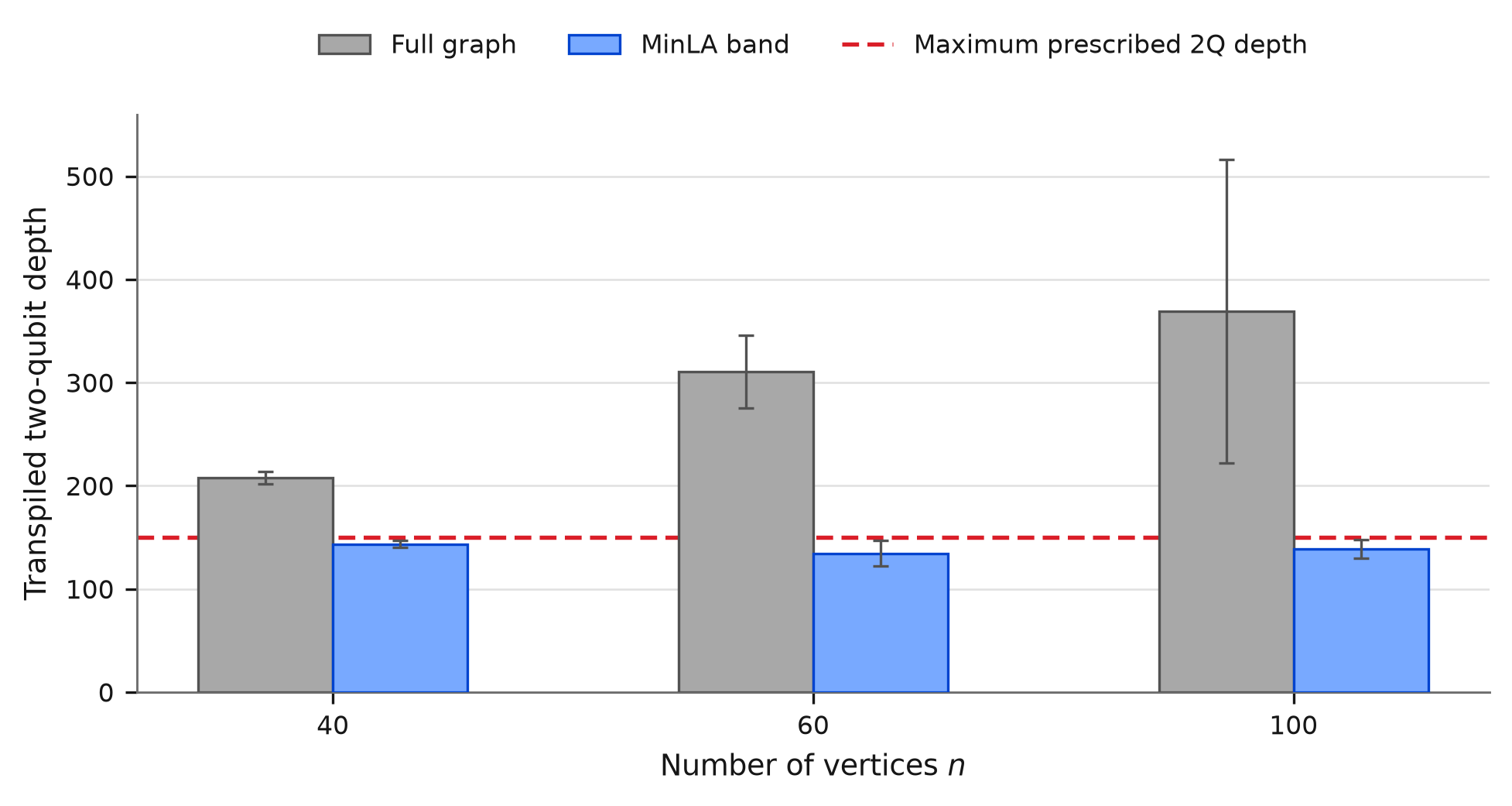}
    \caption{Transpiled two-qubit depth of the full-graph and MinLA-band circuits on \texttt{ibm\_kingston}, paired by instance and compiled at the same $p$. Bars are means over 20 instances; error bars are sample standard deviations. }
\label{fig:regular6-depth}
\end{figure}

In Figure~\ref{fig:regular6-depth}, we report the resulting transpiled two-qubit depths. Despite retaining approximately half of the original interactions at the larger sizes, every surrogate circuit satisfies the hardware depth budget. Across all instances, the mean depth reduction is $31.0\%$, $55.8\%$, and $62.3\%$ at $n=40$, $60$, and $100$, respectively. For details on reduction in the transpiled two-qubit gate counts, readers can refer Table \ref{tab:gate-count-reduction} in Appendix \ref{app:transpiled-two-qubit-gate-counts}.

These results validate the surrogate-construction strategy developed in Section~\ref{sec:interaction-filtering}. Rather than aggressively reducing the graph to a chain-like structure, MinLA ordering concentrates interactions so that a substantially richer surrogate Hamiltonian remains executable within the available hardware depth budget.

\begin{figure}[htbp]
    \centering
    \includegraphics[width=0.8\linewidth]{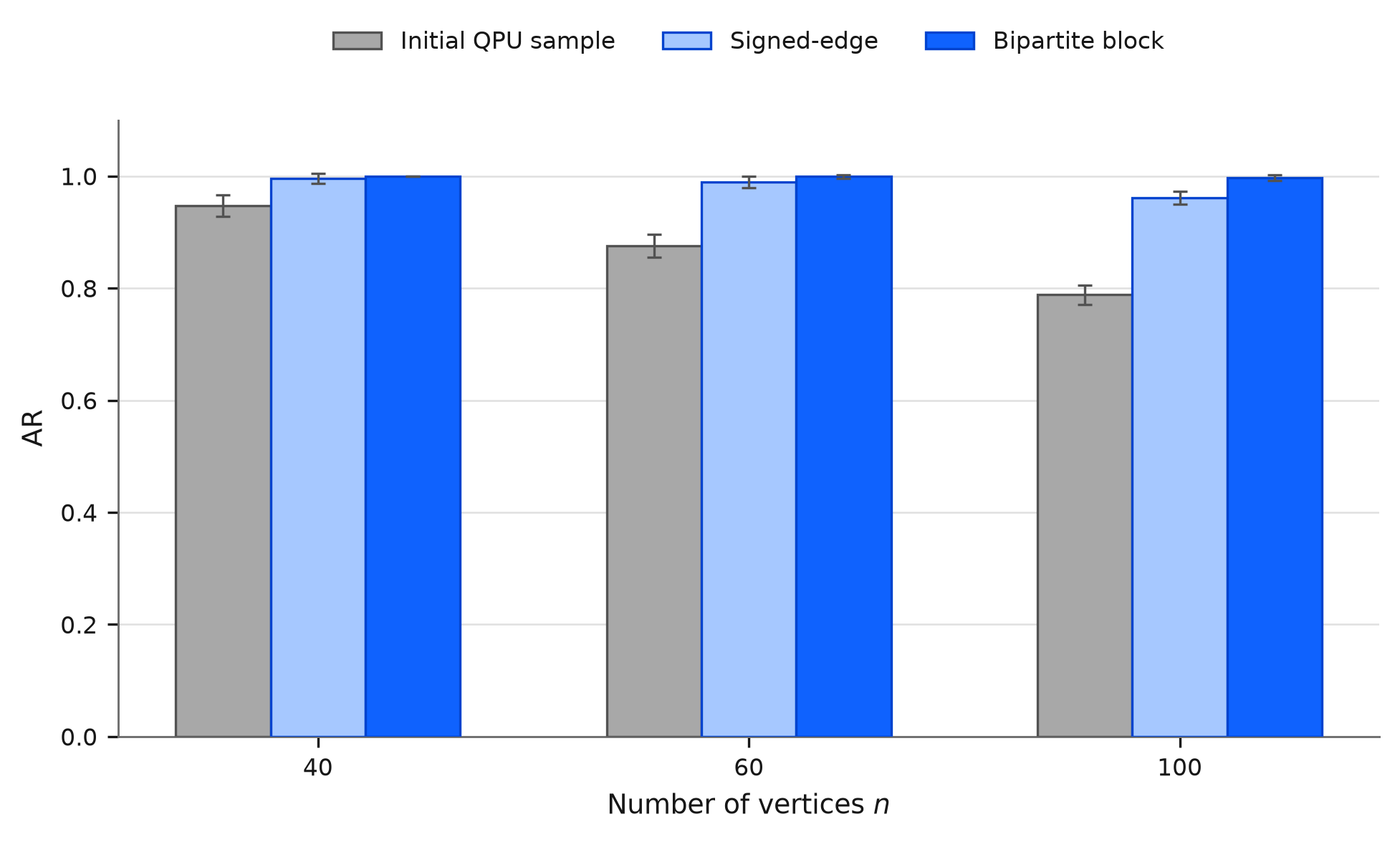}
    \caption{Solution quality on the original graphs, over 20 instances per size. The mean AR before and after recovery, computed per instance from the same 50 MinLA-circuit samples; error bars are sample standard deviations.}
\label{fig:regular6-ar}
\end{figure}

\paragraph{Solution Quality and Classical Recovery:} We next examine the effectiveness of classical recovery in improving configurations sampled from the surrogate QAOA circuits. Figure~\ref{fig:regular6-ar} compares the
best-of-50 approximation ratios before and
after recovery, with all cut values evaluated
on the original graphs. Both recovery strategies improve solution quality across the considered problem sizes, with bipartite-block recovery consistently achieving higher approximation ratios than signed-residue recovery. 

At $n=100$, the mean best-of-50 approximation
ratio increases from $0.788$ before recovery
to $0.961$ following signed-residue recovery
and $0.997$ following bipartite-block recovery. Similar improvements are observed at $n=40$ and $n=60$, where the recovered solutions approach the optimal cut values. These findings demonstrate that configurations obtained from surrogate QAOA circuits can be effectively refined against the original \textsc{MAX-CUT} objective, despite the interactions omitted during quantum execution.


\begin{figure*}[htbp]
    \centering
    \begin{subfigure}[t]{0.48\textwidth}
        \centering
        \includegraphics[width=\linewidth]{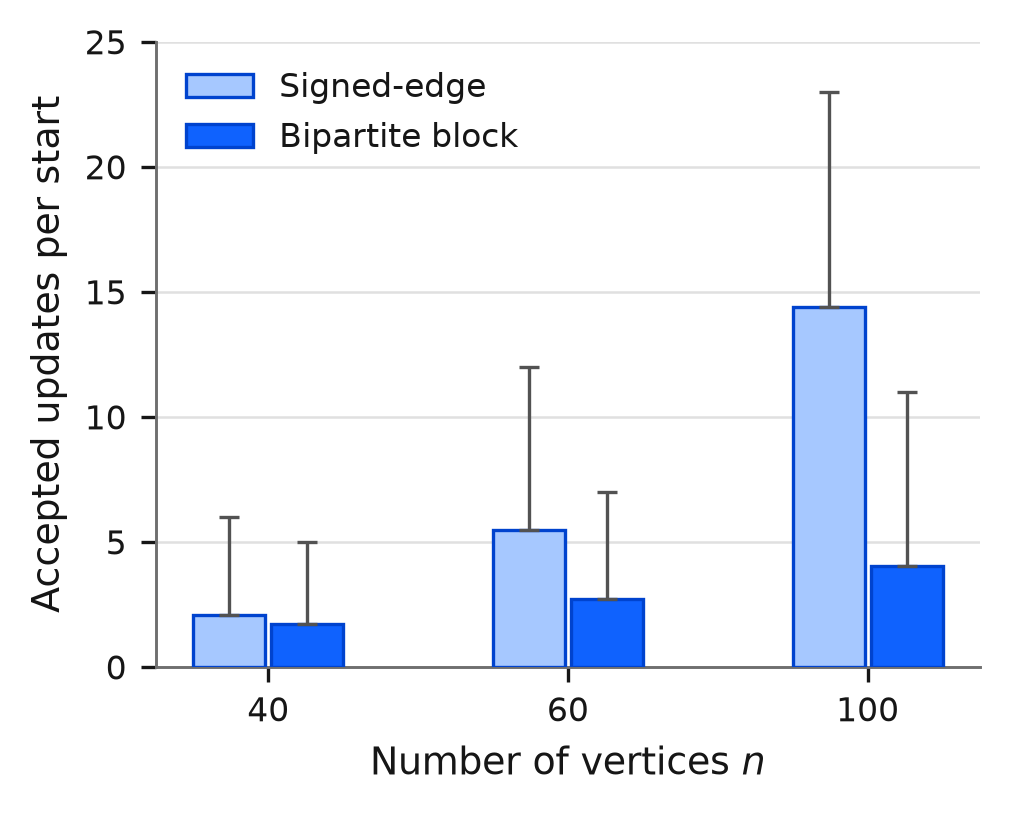}
        \caption{Mean accepted updates per start; whiskers extend to the maximum.}
        \label{fig:regular6-recovery-steps}
    \end{subfigure}
    \hfill
    \begin{subfigure}[t]{0.48\textwidth}
        \centering
        \includegraphics[width=\linewidth]{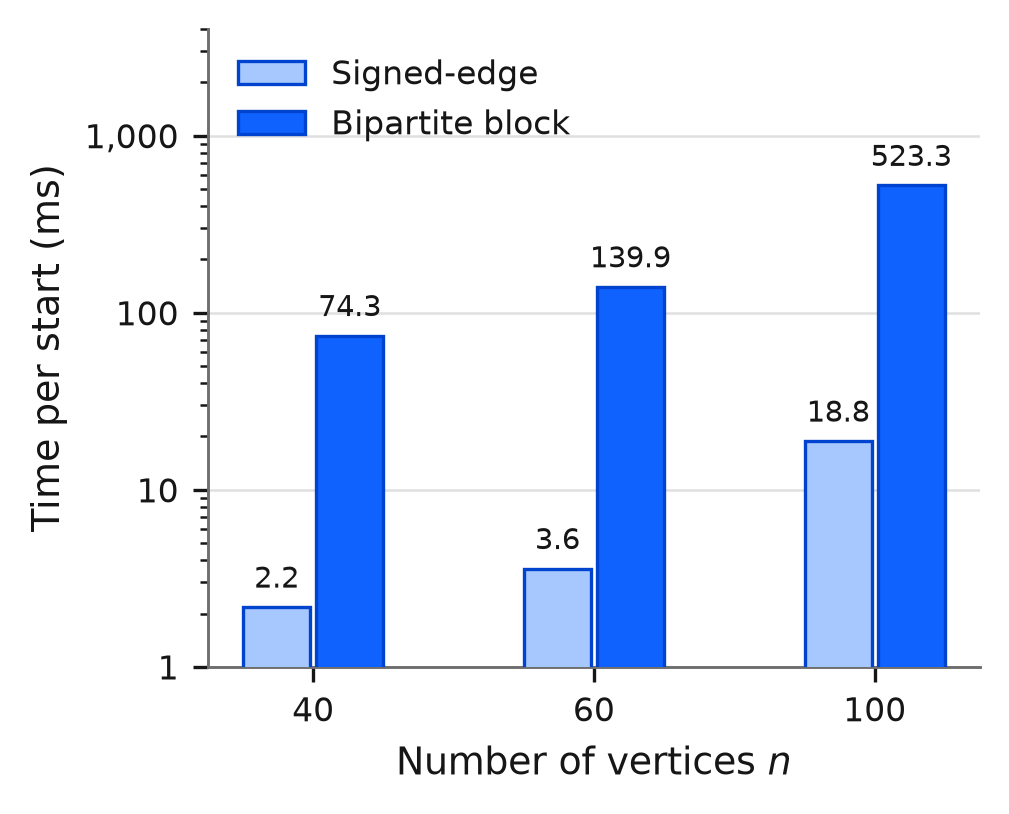}
        \caption{Mean wall time per recovery run, on a logarithmic axis.}
        \label{fig:regular6-recovery-time}
    \end{subfigure}
    \caption{Classical recovery effort over 1,000 runs per method and size: 50 starting configurations for each of 20 instances.}
    \label{fig:regular6-recovery}
\end{figure*}

\paragraph{Classical Recovery Overhead:}
Figure~\ref{fig:regular6-recovery} compares the computational effort of the two recovery
strategies in terms of accepted updates and
execution time. Signed-residue recovery performs inexpensive local updates and therefore converges quickly, whereas bipartite-block recovery repeatedly solves conditional minimum-cut subproblems that explore larger neighborhoods of the solution space. Consequently, bipartite-block recovery requires more computation per recovery run. At $n=100$, the average runtime is 523\,ms compared with 19\,ms for signed-residue recovery. This additional computation is performed entirely classically and requires no further QPU executions. Moreover, recovery runs are independent across starting configurations and can therefore be parallelized.


Taken together, these results support the central hypothesis of this work. The 6-regular experiments demonstrate that hardware-aware surrogate construction can reduce circuit depth while retaining sufficient graph structure to support effective classical recovery. At $n=100$, the complete workflow achieves a mean best-of-50 approximation ratio of $0.997$ on the original graph, without additional quantum executions for recovery.


\subsection{Results on Erd\H{o}s--R\'enyi Graphs}
\label{subsec:gnp-results}

We evaluate the workflow on the HAMLIB \texttt{gnp-k\_4} family, using 20 instances at each size $n\in\{40,60,100\}$. These graphs have non-uniform realized degrees and more edges than the corresponding 6-regular instances. Their mean degrees are approximately 14.1, 16.1, and 18.2, respectively. For each graph, we use $p=1$ and train the angles using a $15\times15$ parameter scan followed by classical refinement \cite{qaoa-training-pipeline,Egger_2021}. The parameters are trained once per graph and reused for its full and MinLA-band circuits. Mean training times are 9.42, 19.23, and 50.67 seconds per graph at $n=40$, $60$, and $100$, respectively, and is completely classical.

\paragraph{Surrogate Construction and Circuit Depth:} We prescribe a maximum transpiled two-qubit depth of $L=200$, as mentioned in Table \ref{tab:exp-setup}.
%
%
Previous experiments on IBM superconducting processors have demonstrated meaningful results at comparable or greater two-qubit circuit  depths~\cite{robledo2025chemistry,Barroca_2026}. Table~\ref{tab:gnp-minla} summarizes the selected bands. At $n=40$ and $n=60$,
the full-graph circuits satisfy the depth
constraint, allowing all interactions to be
retained. At $n=100$, the selected bands retain $52.2\%$ of the original edges on average.

\begin{table}[htbp]
\centering
\caption{Selected MinLA bands for 20 HAMLIB \texttt{gnp-k\_4} instances
per size under $L=200$. The $p^\star$ column counts instances selected
at each QAOA depth. Entries are means $\pm$ sample standard deviations;
ranges are given in brackets.}
\label{tab:gnp-minla}
\footnotesize
\setlength{\tabcolsep}{3pt}
\renewcommand{\arraystretch}{1.2}
\begin{tabular}{|r|c|c|c|c|c|}
\hline
$n$ & $\overline{|E|}$ & $\overline{C_G^\star}$ &
$p^\star=1/2/3$ & $\overline{k^\star}$ [range] & Retained (\%) \\
\hline
40  & $283.0\pm10.7$ & $186.3\pm6.8$ & $20/0/0$
    & $36.2\pm1.5$ $[33,39]$ & $100.0\pm0.0$ \\
60  & $483.9\pm17.4$ & $318.9\pm10.3$ & $20/0/0$
    & $55.1\pm1.9$ $[51,58]$ & $100.0\pm0.0$ \\
100 & $910.7\pm32.5$ & $600.1\pm18.1$ & $20/0/0$
    & $19.0\pm1.2$ $[16,21]$ & $52.2\pm2.6$ \\
\hline
\end{tabular}
\end{table}

\begin{figure}[htbp]
    \centering
    \includegraphics[width=0.8\linewidth]{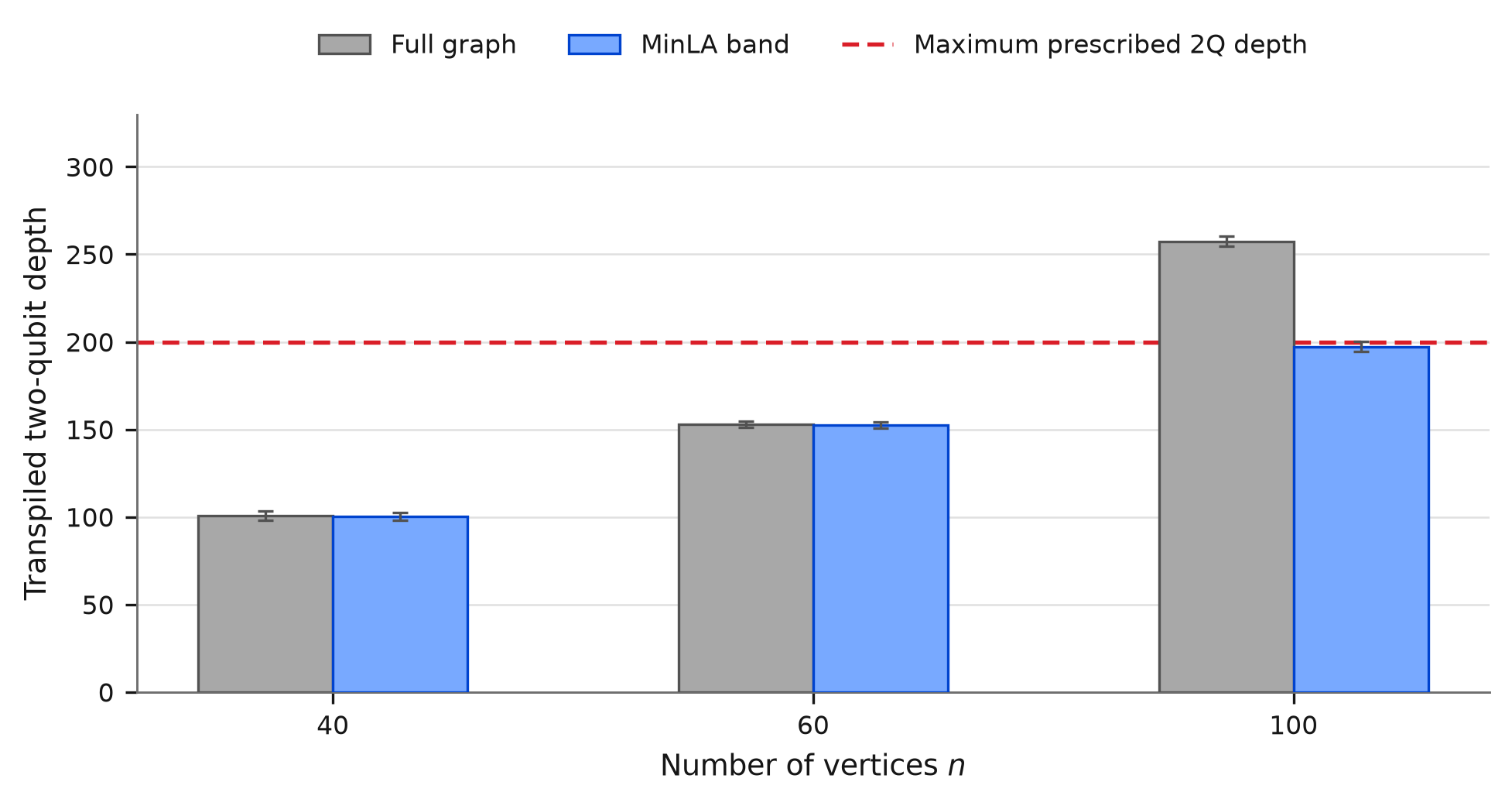}
    \caption{Transpiled two-qubit depth of the full-graph and MinLA-band
    circuits on \texttt{ibm\_kingston}, paired by instance. Bars are
    means over 20 instances; error bars are sample standard deviations.
    The red dashed line marks the maximum prescribed depth $L=200$.}
    \label{fig:gnp-depth}
\end{figure}

Figure~\ref{fig:gnp-depth} reports the transpiled two-qubit depths. At $n=40$ and $n=60$, the depths remain nearly unchanged, consistent with full interaction retention. We emphasise that full-edge retention at $n=40$ and $n=60$ is specific to the tested $p=1$ circuits. While we have not tested these instances for $p>1$, increasing $p$ pushes the full-graph $(n=40,60)$ circuits' two-qubit depths above $200$, necessitating a narrower MinLA band and dropping some edges. At $n=100$, the band retains $52.2\%$ of the edges and reduces mean two-qubit depth from 257.4 to 197.4, a paired mean reduction of
$23.3\%$. The mean two-qubit gate count also decreases from 12,379 to 9,225, a reduction of $25.5\%$. 

\paragraph{Solution Quality and Classical Recovery:}
We next evaluate whether configurations obtained from the QAOA circuits can be refined to produce high-quality solutions to the original \textsc{MAX-CUT} instances.

\begin{figure}[htbp]
    \centering
    \includegraphics[width=0.8\linewidth]{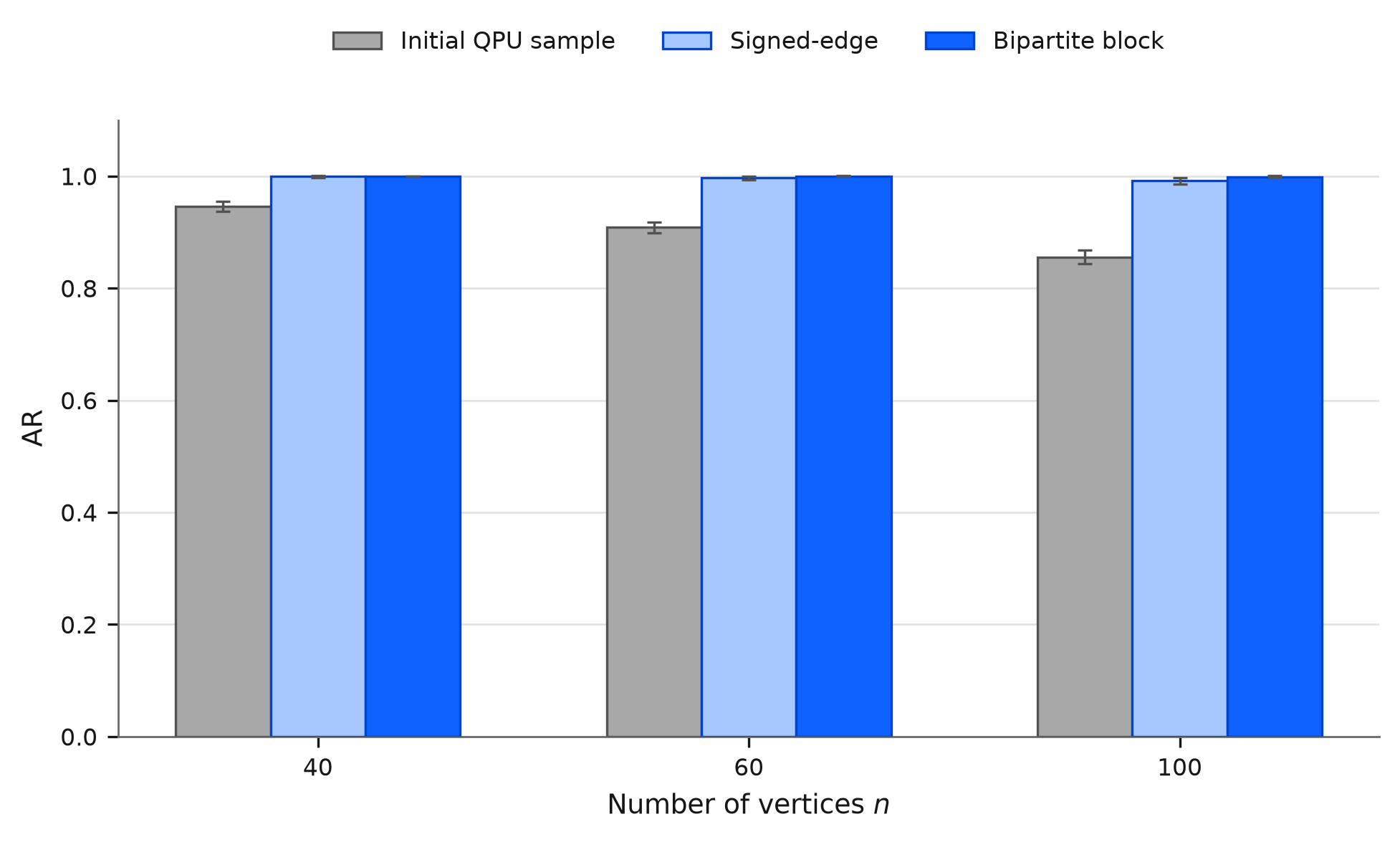}
    \caption{Best-of-50 approximation ratios on the original $G(n,p)$
    graphs, over 20 instances per size for $p=1$ QAOA cost layers. The initial QPU sample and both recovered solutions use the same 50 starting configurations per
    instance. Error bars are sample standard deviations.}
    \label{fig:gnp-ar}
\end{figure}

Figure~\ref{fig:gnp-ar} compares the best-of-50 approximation ratios before and after recovery, with all cut values evaluated on the original graphs. Both recovery strategies improve the sampled configurations, with bipartite-block recovery achieving higher mean approximation ratios than signed-residue recovery. For $n=40$ and $n=60$, where all interactions are retained, these improvements reflect classical refinement of full-graph QAOA samples. At $n=100$, recovery additionally addresses the interactions omitted during surrogate construction, improving the resulting configurations against the complete objective.

\paragraph{Classical Recovery Overhead:}

\begin{figure*}[htbp]
    \centering
    \begin{subfigure}[t]{0.48\textwidth}
        \centering
        \includegraphics[width=\linewidth]{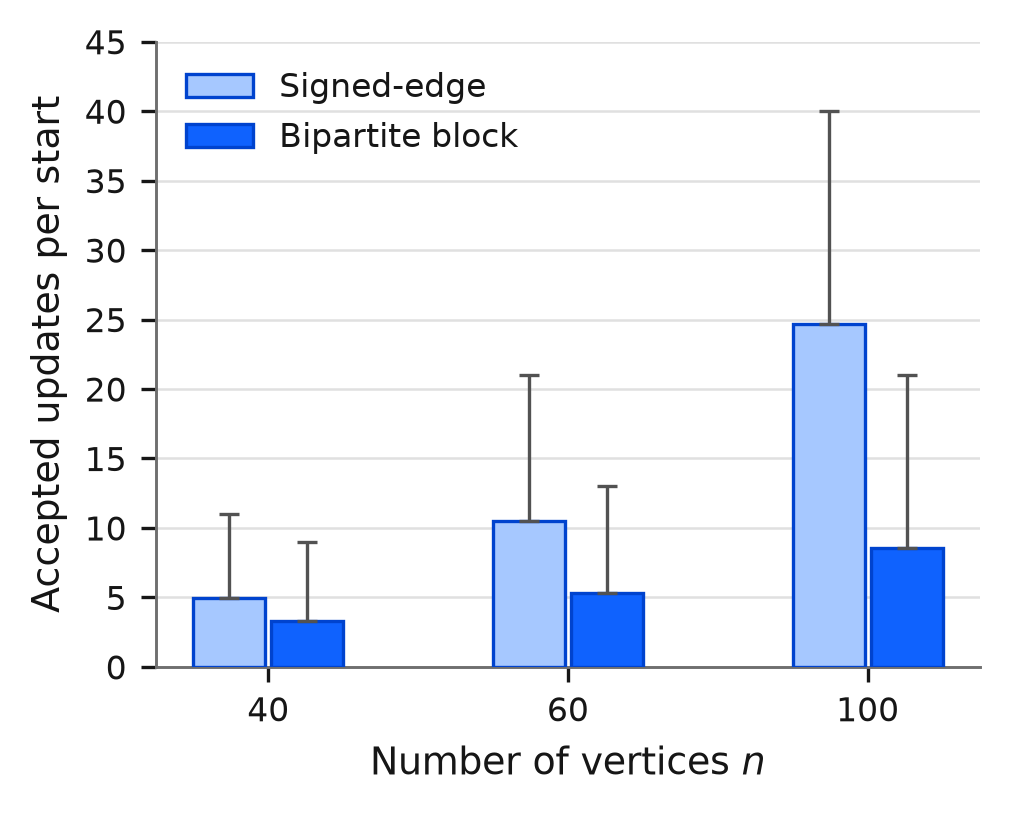}
        \caption{Mean accepted updates per start; whiskers extend to the
        maximum.}
        \label{fig:gnp-recovery-steps}
    \end{subfigure}
    \hfill
    \begin{subfigure}[t]{0.48\textwidth}
        \centering
        \includegraphics[width=\linewidth]{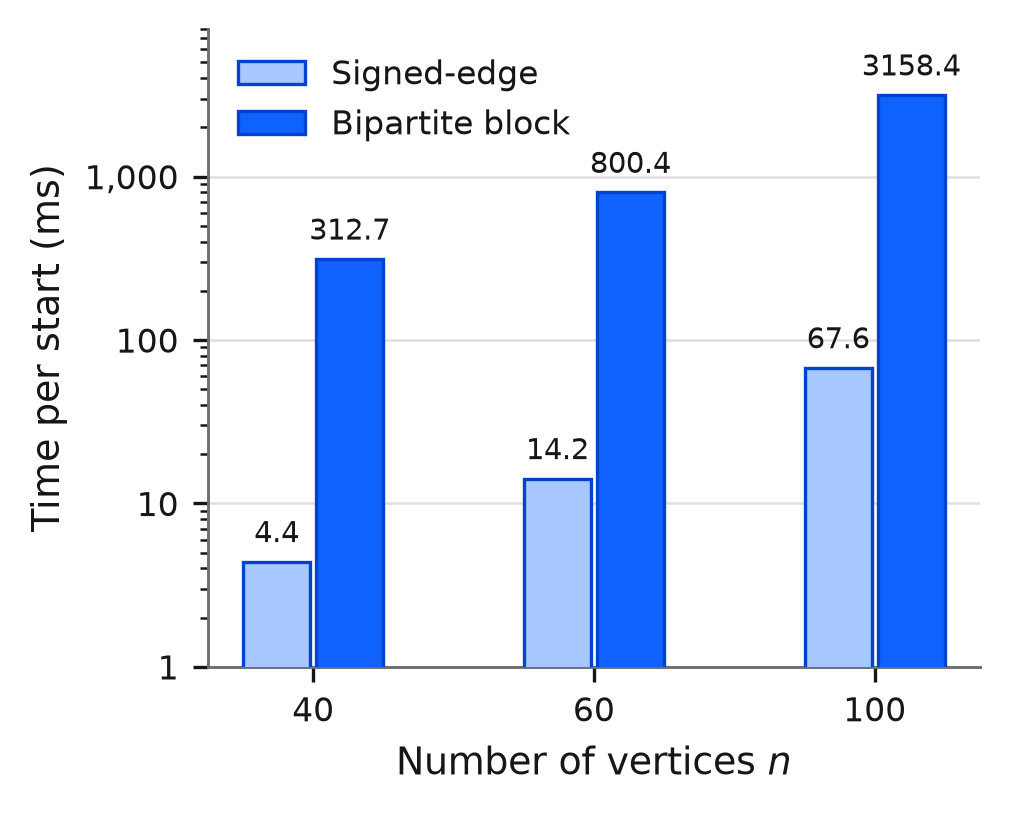}
        \caption{Mean wall time per recovery start, on a logarithmic
        axis.}
        \label{fig:gnp-recovery-time}
    \end{subfigure}
    \caption{Classical recovery effort over 1,000 starts per method and
    size: 50 starting configurations for each of 20 instances.}
    \label{fig:gnp-recovery}
\end{figure*}

Figure~\ref{fig:gnp-recovery} compares the
computational effort of signed-residue
and bipartite-block recovery in terms
of accepted updates and execution time. As observed for the 6-regular instances,
bipartite-block recovery achieves higher
solution quality at the expense of greater
classical computation. 
%
At $n=100$, signed-edge recovery takes 67.6\,ms per start, while bipartite-block recovery takes 3.16\,s. These statistics are computed over 1,000 recovery runs per method and size, corresponding to 50 starting configurations for each of 20 graph instances. Since recovery operates independently on each configuration, the computation was parallelized without additional QPU executions.

Overall, these experiments demonstrate
that the need for interaction reduction
depends on graph size and circuit depth.
While the full $p=1$ Hamiltonian remains
executable within the prescribed budget
at smaller sizes, depth-constrained
surrogates become necessary at $n=100$.
Classical recovery further improves the
sampled configurations, without increasing
the quantum execution cost.

\subsection{Cross-Family Comparison}
\label{subsec:cross-family}

Table~\ref{tab:cross-family-retention} compares the edge-retention results under the family-specific depth limits. We chose the limits to target retention of approximately half of the edges in the largest instances; this led to $L=150$ for the 6-regular experiments and $L=200$ for $G(n,p)$. At $n=100$, both families retain approximately half of their interactions. At smaller sizes, the $G(n,p)$ bands retain all edges under $L=200$, since the full $p=1$ circuits already satisfy that limit. The 6-regular experiments use greater QAOA depths for many instances, and consequently require edge removal to meet $L=150$.

\begin{table}[htbp]
\centering
\caption{Edge retention and QAOA settings across the two graph families.
The depth limits are family-specific.}
\label{tab:cross-family-retention}
\footnotesize
\setlength{\tabcolsep}{4pt}
\renewcommand{\arraystretch}{1.15}
\begin{tabular}{|r|c|c|c|c|c|}
\hline
$n$ & 6-regular $p^\star=1/2/3$ & $L$ & Retained (\%) &
$G(n,p)$ $p$ & $L$ / Retained (\%) \\
\hline
40  & $0/0/20$ & 150 & $70.5\pm3.2$  & 1 & $200 / 100.0$ \\
60  & $0/2/18$ & 150 & $49.2\pm5.8$  & 1 & $200 / 100.0$ \\
100 & $6/8/6$  & 150 & $51.4\pm15.7$ & 1 & $200 / 52.2\pm2.6$ \\
\hline
\end{tabular}
\end{table}

This comparison describes two operating regimes, rather than isolating
the effect of graph family: QAOA depth and the prescribed depth limit both differ. The observed retention therefore reflects graph structure, QAOA depth, and the compilation constraint together; it should not be
interpreted as evidence that either family is intrinsically easier to
compile. Likewise, the approximation ratios are evaluated within each
family's experiment and are not a controlled comparison of $p$ or
topology.

\section{Discussion}
\label{sec:discuss}

In this section, we highlight the scope of our experimental findings, particularly how hardware-aware circuit reduction and classical recovery contribute to the proposed workflow, and its adaptability to future quantum hardware.

\begin{itemize}
    \item \textit{Hardware-aware execution under noise constraints}: The primary contribution of our work is a systematic approach to executing non-hardware-native QAOA instances within prescribed circuit-depth budgets. By heuristically selecting interactions, our framework balances the retention of graph structure against the routing overhead of quantum execution. This provides a practical means of limiting exposure to hardware noise, although circuit-depth reduction does not necessarily translate directly into improved fidelity. The chosen depth budgets represent empirical operating choices rather than universal hardware thresholds.

    \item \textit{Interpreting solution quality:} The reported approximation ratios characterize the performance of the complete hybrid workflow, comprising surrogate-based quantum sampling and classical configuration recovery. While the quantum circuit generates candidate configurations using a subset of the original interactions, classical recovery improves these configurations against the full objective. The near-optimal recovered cuts therefore establish the effectiveness of the combined procedure, without attributing the final solution quality exclusively to quantum execution. Importantly, the framework preserves the original optimization objective throughout classical recovery, even though the executed Hamiltonian is modified.

    \item \textit{Scalability and Hardware Adaptability:} Our framework separates quantum circuit complexity from classical solution refinement, allowing the computational workload to be adjusted according to available hardware resources. As processors support deeper and more reliable circuits, larger subsets of the original interactions can be retained without modifying the overall workflow. Furthermore, independent recovery across sampled configurations enables parallel classical processing without additional QPU executions. The experiments establish the feasibility of this approach for instances with up to 100 vertices, while its performance at larger scales will depend on both quantum execution requirements and classical recovery costs.
\end{itemize}

\section{Conclusion}
\label{sec:conclusion}

Sparse connectivity and noise remain major obstacles to executing QAOA on graph instances whose interaction structure differs substantially from the coupling topology of current superconducting quantum processors. In this work, we introduced a hybrid quantum-classical workflow that addresses this challenge through depth-constrained surrogate Hamiltonian construction and classical configuration recovery. By combining MinLA-based band-concentrating vertex orderings with hardware-aware interaction filtering, we construct surrogate Hamiltonians that retain all decision variables and a substantial fraction of the original graph structure while satisfying a prescribed two-qubit depth budget. We further developed signed-residue and bipartite-block recovery procedures that operate on the original graph and monotonically improve sampled solutions without requiring additional QPU executions.

Using this workflow, we demonstrated QAOA on non-hardware-native 6-regular and Erd\H{o}s--R\'enyi graph instances with up to 100 vertices on the \texttt{ibm\_kingston} processor. Although the full Hamiltonians of these problems incur substantial routing overhead when mapped directly to sparse superconducting hardware, the proposed surrogate circuits remained executable within realistic depth constraints and, when combined with configuration recovery, achieved approximation ratios approaching unity on the original graph. These results show that useful optimization signals can be extracted from hardware-feasible surrogate Hamiltonians and subsequently amplified through recovery on the complete objective.

More broadly, the proposed workflow provides a practical pathway toward scalability on near-term quantum hardware. The depth budget used to construct the surrogate Hamiltonian is fundamentally determined by the level of noise that current devices can tolerate. As superconducting processors continue to improve, deeper circuits will become executable, allowing larger interaction radii and therefore richer surrogate Hamiltonians that preserve an even greater fraction of the original graph. The resulting increase in retained problem information can strengthen the optimization signal available to the quantum processor, reducing the burden on subsequent recovery and extending the range of graph instances that can be addressed. In this sense, surrogate-based execution and configuration recovery offer a scalable framework for bringing increasingly large and non-hardware-native quantum optimization problems within reach of future superconducting quantum computers.

\section*{Acknowledgments}
The authors acknowledge helpful discussions with Dr. Shesha Raghunathan, Dr. Kalyan Dasgupta and Dr. Sumanta Mukherjee. RM acknowledges helpful discussions with Dr. Nick Bronn.

\bibliographystyle{unsrt}  
\bibliography{references}  

\appendix

\section{Additional Preliminaries}
\label{app:prelim}

\subsection{QAOA Objective and Circuit Decomposition}
\label{app:qaoa-details}

The cost Hamiltonian in
Eq.~\eqref{eq:maxcut-ising} satisfies
\begin{equation}
\langle\boldsymbol{x}|H_C|
\boldsymbol{x}\rangle
=C_G(\boldsymbol{x}).
\end{equation}
Therefore, the ground states of $-H_C$
correspond to maximum cuts of $G$.

The QAOA parameters are chosen to minimize
the expected energy,
\begin{equation}
\label{eq:qaoa-objective}
E_p(\boldsymbol{\theta})
=
\langle\psi_p(\boldsymbol{\theta})|
-H_C|\psi_p(\boldsymbol{\theta})\rangle.
\end{equation}
For fixed parameters, measurements in the
computational basis yield candidate bitstrings
with probabilities
\begin{equation}
P_{\boldsymbol{\theta}}(\boldsymbol{x})
=
|\langle\boldsymbol{x}|
\psi_p(\boldsymbol{\theta})\rangle|^2,
\end{equation}
from which the best observed partition can
be selected.

Since the terms of $H_C$ commute, each
cost layer decomposes, up to a global phase, as
\begin{equation}
\label{eq:cost-decomposition-app}
U_C(\gamma)\doteq
\prod_{(u,v)\in E}
R_{ZZ}^{(u,v)}(-\gamma),
\end{equation}
where
$R_{ZZ}(\phi)=
e^{-i\phi Z\otimes Z/2}$.
Thus, a $p$-layer QAOA circuit contains
$p|E|$ logical two-qubit interactions,
whose scheduling and hardware implementation
determine the resulting circuit depth.

\subsection{QAOA Parameter Selection}
\label{subsec:angle-selection}

The standard variational approach determines the QAOA parameters through classical optimization,
    \begin{equation} \label{eq:angle-optimization}
        \boldsymbol{\theta}^{\star} =\arg\min_{\boldsymbol{\theta}}
        E_p(\boldsymbol{\theta}).
    \end{equation}
Evaluating this objective on quantum hardware
requires repeated circuit executions, making
the optimization susceptible to well-known challenges of variational quantum algorithms, including shot
noise, device noise, optimization instability, and barren plateaus~\citep{mcclean2018barren, cerezo2021variational}. Consequently, parameter
selection methods that exploit graph structure
and avoid repeated hardware-based optimization
are of practical interest ~\cite{qaoa-training-pipeline}. 
A useful property of QAOA is the locality of its expectation values. At depth $p$, the expectation of an edge observable $Z_uZ_v$ depends only on the subgraph within its $p$-hop light cone~\citep{farhi2014quantum}.
This allows parameters to be transferred between graphs with similar local structure and motivates graph-family-specific angle selection methods~\citep{brandao2018fixed,
galda2021transferabilityoptimalqaoaparameters}.

\subsubsection{Regular instances} 
For a $d$-regular graph, we use the fixed-angle approach of \citep{fixedangle2021wurtz}, which exploits QAOA locality to obtain parameters from the corresponding infinite $d$-regular tree
    \begin{equation}
        (\boldsymbol{\gamma}^{\mathrm{FA}},
         \boldsymbol{\beta}^{\mathrm{FA}})
        =
        \mathcal{A}_{\mathrm{FA}}(d,p).
    \end{equation}
For $d$-regular graphs with girth greater than $2p+1$, all edge light cones have the same tree structure. The expected cut fraction therefore depends only on $d$, $p$, the QAOA parameters, and not on the graph size or specific instance. For regular graphs containing short cycles, these angles may also serve as transferable heuristic parameters, without requiring instance-specific quantum optimization.


\subsubsection{Erd\H{o}s--R\'enyi instances.}
For irregular graphs, the variation in vertex degrees and local connectivity precludes a single regular-tree description. We therefore obtain instance-specific parameters at $p=1$ using the graph-training methodology of the QAOA training pipeline~\citep{qaoa-training-pipeline}. This method performs a coarse $(\beta,\gamma)$ grid search followed by local refinement. The \texttt{EfficientDepthOneEvaluator} computes the required correlators from reduced $4\times4$ density matrices, avoiding full statevector simulation and remaining practical for larger instances~\citep{Egger_2021}.

These procedures provide all variational parameters used in the subsequent hardware experiments without requiring iterative QPU-based optimization.

\subsection{Connectivity-Aware Compilation of QAOA Circuits}
\label{subsec:circuit-compilation}

QAOA implementation on hardware is constrained by the limited connectivity between physical qubits. Although the commuting two-qubit interactions within each cost layer can be reordered, their execution requires suitable qubit placement, routing, and gate scheduling. Consequently, the compiled circuit depth
depends not only on the structure of the problem graph
but also on how its interactions are mapped onto the
hardware topology. We review three very useful
compilation techniques: (a) SWAP strategies for routing two-qubit interactions, (b) \texttt{SATMapper} for optimizing the initial qubit placement, and (c) edge coloring for exploiting parallelism among commuting gates.

\paragraph{\textbf{SWAP strategy}:} On sparsely connected hardware, two-qubit interactions
can be executed only between adjacent physical qubits.
The line-SWAP strategy~\citep{Weidenfeller2022}
addresses this constraint by routing logical qubits
along a path $(q_0,\ldots,q_{n-1})$ embedded in the
hardware coupling graph. It alternates between
two layers of disjoint SWAP operations,
    \begin{equation} \label{eq:swap-layers}
        \begin{aligned}
        \mathcal{S}_{\mathrm{even}}
        &=\{(q_0,q_1),(q_2,q_3),\ldots\},\\
        \mathcal{S}_{\mathrm{odd}}
        &=\{(q_1,q_2),(q_3,q_4),\ldots\}.
        \end{aligned}
    \end{equation}
These alternating layers systematically change the
positions of logical qubits, allowing the required
$R_{ZZ}$ interactions to be applied when their
corresponding qubits become adjacent. Since the
cost-Hamiltonian terms commute, the interactions
can be reordered to follow this routing sequence.

\paragraph{\textbf{SATMapper}:}The number of SWAP layers required by a given strategy depends on the initial assignment of logical qubits to physical qubits.
\texttt{SATMapper}~\citep{MATSUO_2023} formulates this
assignment as a Boolean satisfiability problem.
For an initial mapping $\mu$, let
$\mathcal{A}_L(\mu)$ denote the set of logical
qubit pairs that become adjacent within the
first $L$ layers of the SWAP strategy.
The objective is to determine
    \begin{equation}  \label{eq:sat-mapping}
        L^\star=
        \min_{\mu,L}
        \left\{
        L:E(G)\subseteq\mathcal{A}_L(\mu)
        \right\}.
    \end{equation}
The \texttt{SATMapper} searches for a feasible initial mapping while minimizing the required number of SWAP
layers. By placing frequently interacting logical
qubits advantageously along the routing path,
it substantially reduces the routing overhead
relative to arbitrary initial placements.

\paragraph{Edge coloring and gate scheduling:} The commuting $R_{ZZ}$ interactions can be scheduled in parallel whenever they act on disjoint qubit pairs. This scheduling problem is naturally represented by an \textit{edge coloring} of the interaction graph, where edges sharing a vertex receive different colors. Each color class therefore defines a set of mutually
disjoint interactions. In practice, greedy edge coloring provides an efficient ordering of the $R_{ZZ}$ gates \cite{majumdar2021optimizing}.
With all these techniques combined, this ordering helps reduce the overhead of implementing QAOA cost layers on connectivity-constrained hardware while preserving the interactions
specified by the problem Hamiltonian.

\section{Detailed Proofs}
\subsection{Proof of Lemma \ref{lem:residue-update}}
\label{sec:proof-of-lem-res-update}

\begin{proof}
Let $i$ be the flipped vertex. This gives that $x_i'=1-x_i$ and $x_j'=x_j$ for every $j\neq i$. We distinguish three cases.

\textit{Case $j=i$:} Using the definition of the signed vertex residue \eqref{eq:vertex-residue}, we have,
\begin{equation}
    \rho_i(\mathbf x)
    =
    \sum_{j\in N_G(i)}
    \bigl(1-2\mathbf 1[x_i\neq x_j]\bigr).
\end{equation}
Every edge incident on $i$ satisfies
\begin{equation}
    \mathbf 1[x_i'\neq x_j']
    =
    1-\mathbf 1[x_i\neq x_j].
\end{equation}
The above equation can be interpreted as if $(i,j)$ is cut after flipping $i$, then it was originally a bad/uncut edge. Consequently,
\begin{equation}
    \rho_i(\mathbf x')
    =
    \sum_{j\in N_G(i)}
    \bigl(1-2(1-\mathbf 1[x_i\neq x_j])\bigr)
    =
    -\rho_i(\mathbf x).
\end{equation}

\textit{Case $j\in N_G(i)$:} Only the contribution of edge $(i,j)$ to $\rho_j$ changes. Writing $a_{ij}=\mathbf 1[x_i\neq x_j]$, we obtain
\begin{equation}
    \rho_j(\mathbf x')-\rho_j(\mathbf x)
    =
    \bigl(1-2(1-a_{ij})\bigr)
    -
    \bigl(1-2a_{ij}\bigr)
    =
    -2(1-2a_{ij}).
\end{equation}

\textit{Otherwise:} Neither the assignment of $j$ nor that of any of its neighbors changes. Every contribution to its residue therefore remains unchanged, giving
\begin{equation}
    \rho_j(\mathbf x')=\rho_j(\mathbf x).
\end{equation}
\end{proof}

\subsection{Proof of Theorem \ref{thm:greedy-monotonicity}}
\label{sec:proof-of-thm-greedy-mono}

\begin{proof}
If vertex $i_t$ is selected at iteration $t$, the acceptance condition and Eq.~\eqref{eq:vertex-gain} give
\begin{equation}
    C_G(\mathbf x^{(t+1)})
    -
    C_G(\mathbf x^{(t)})
    =
    \rho_{i_t}(\mathbf x^{(t)})>0.
\end{equation}
The final inequality follows by summing the accepted gains, with equality when no flip is accepted.
\end{proof}

\subsection{Proof of Theorem \ref{thm:greedy-termination}}
\label{sec:proof-of-thm-greedy-term}

\begin{proof}
By Theorem~\ref{thm:greedy-monotonicity},
every accepted flip strictly increases the
cut value. Since $C_G$ is integer-valued,
each improvement is at least one. Moreover,
$C_G(\mathbf{x})\leq |E|$ for every configuration.
Consequently, after $F$ accepted flips,
    \begin{equation}
        C_G(\mathbf{x}^{(0)})+F
        \leq C_G(\mathbf{x}^{(F)})\leq |E|,
    \end{equation}
which establishes the bound in Eq.~\eqref{eq:flip-bound}. By Eq.~\eqref{eq:vertex-gain}, the residue
$\rho_i(\mathbf{x})$ equals the change in cut
value obtained by flipping vertex $i$.
Therefore, if all residues are nonpositive,
no single-vertex flip can improve the
objective, establishing one-flip local
optimality.
\end{proof}

\subsection{Proof of Corollary \ref{cor:one-flip-guarantee}}
\label{sec:proof-of-cor-sing-flip}

\begin{proof}
At a one-flip local optimum, no single-vertex
flip can improve the cut. Consequently,
$\rho_i(\widehat{\mathbf{x}})\leq 0$ for every
$i\in V$, implying
\begin{equation}
\sum_{i\in V} B_i(\widehat{\mathbf{x}})
\leq
\sum_{i\in V} G_i(\widehat{\mathbf{x}}).
\end{equation}
Each uncut edge contributes once to the
uncut-edge count of each endpoint, while
each cut edge contributes similarly to
the cut-edge counts. Therefore,
\begin{equation}
\begin{aligned}
\sum_{i\in V}B_i(\widehat{\mathbf{x}})
=2\bigl(|E|-C_G(\widehat{\mathbf{x}})\bigr), \quad 
\sum_{i\in V}G_i(\widehat{\mathbf{x}})
=2C_G(\widehat{\mathbf{x}}).
\end{aligned}
\end{equation}
It follows that $2\bigl(|E|-C_G(\widehat{\mathbf{x}})\bigr)
\leq 2C_G(\widehat{\mathbf{x}})$ yielding
$C_G(\widehat{\mathbf{x}})\geq |E|/2$.
Since $C_G^\star\leq |E|$, the claimed
approximation bound follows.
\end{proof}

\section{Transpiled Two-Qubit Gate Counts}
\label{app:transpiled-two-qubit-gate-counts}

To quantify the circuit-resource savings from graph filtering, we compare each selected surrogate with the corresponding full-graph
circuit at the same QAOA depth. Table~\ref{tab:gate-count-reduction} reports the mean transpiled two-qubit gate counts over 20 instances per graph family and size. All two-qubit gates in the recorded compiled circuits are $\operatorname{CZ}$ gates.

\begin{table}[htbp]
\centering
\caption{Transpiled two-qubit gate counts for full-graph and selected
surrogate circuits at matched QAOA depth. Counts are averaged over
20 instances. The mean reduction is computed as
$\frac{100}{20}\sum_{i=1}^{20}
(1-N_{\mathrm{sur},i}/N_{\mathrm{full},i})$,
where $N_{\mathrm{full},i}$ and $N_{\mathrm{sur},i}$ denote the
corresponding two-qubit gate counts.}
\label{tab:gate-count-reduction}
\begin{tabular}{lrrrr}
\hline
Family & $n$ & Full graph & Surrogate & Reduction (\%) \\
\hline
6-regular & 40  & 3769.65  & 2481.15 & 34.12 \\
6-regular & 60  & 8642.85  & 3402.15 & 59.72 \\
6-regular & 100 & 17490.65 & 6004.95 & 57.52 \\
$G(n,p)$       & 40  & 1881.45  & 1876.00 & 0.28  \\
$G(n,p)$        & 60  & 4344.85  & 4343.30 & 0.04  \\
$G(n,p)$        & 100 & 12379.45 & 9224.65 & 25.47 \\
\hline
\end{tabular}
\end{table}

For 6-regular graphs, the mean reductions are $34.1\%$, $59.7\%$, and $57.5\%$ at $n=40,60,100$, respectively. For $G(n,p)$ graphs, the mean reduction is $25.5\%$ at $n=100$. Gate counts remain essentially unchanged at $n=40$ and $n=60$, where the selected surrogates retain all original edges; the small differences arise between the compiled implementations.

\end{document}